\documentclass[floats,floatfix,showpacs,amssymb,prd,twocolumn,superscriptaddress,nofootinbib,longbibliography]{revtex4-1}
\usepackage{anyfontsize}
\usepackage{amssymb}
\usepackage{amsmath}
\usepackage{amsfonts}
\usepackage{amsbsy}
\usepackage[utf8]{inputenc} 
\usepackage{newtxtext}
\usepackage{newtxmath}
\usepackage{tensor}
\usepackage{mathrsfs}
\usepackage{bm}
\usepackage[utf8]{inputenc}
\usepackage{graphicx}
\usepackage{epsfig}
\usepackage{epstopdf}
\usepackage[usenames,dvipsnames]{xcolor}
\usepackage{verbatim,mathtools,needspace,enumitem,etoolbox}%,physics,microtype}

\definecolor{linkcolor}{rgb}{0.0,0.3,0.5}

\usepackage[unicode, colorlinks=true, linkcolor=linkcolor, citecolor=linkcolor, filecolor=linkcolor,urlcolor=linkcolor, pdfusetitle]{hyperref}

\usepackage{epstopdf}

\usepackage{bm}
\usepackage{dcolumn}

\usepackage{longtable}
\usepackage{enumerate}
 \usepackage[normalem]{ulem}

\hypersetup{colorlinks=true,
            citecolor=blue,
            linkcolor=blue,
            urlcolor=blue}
\usepackage[T1]{fontenc}
\usepackage{subfigure}
\usepackage{amsmath}
\usepackage{float}
\usepackage{mathrsfs}
\usepackage[dvipsnames]{xcolor}
\usepackage{soul}

\pdfoutput=1
\begin{document}
\sloppy

\title{On-Axis Tidal Forces in Kerr Spacetime}

\author{Haroldo C. D. Lima Junior}
\email{haroldo.ufpa@gmail.com}
\affiliation{Faculdade de F\'{\i}sica,
Universidade Federal do Par\'a, 66075-110, Bel\'em, PA, Brasil }

\author{Lu\'{\i}s C. B. Crispino}
\email{crispino@ufpa.br}
\affiliation{Faculdade de F\'{\i}sica,
Universidade Federal do Par\'a, 66075-110, Bel\'em, PA, Brasil }

\author{Atsushi Higuchi}
\email{atsushi.higuchi@york.ac.uk}
\affiliation{Department of Mathematics, University of York, Heslington, York YO10 5DD, UK}

\begin{abstract}
Tidal forces are an important feature of General Relativity, which are related to the curvature tensor. 
We analyze the tidal tensor in Kerr spacetime, with emphasis on the case along the symmetry axis of the Kerr black hole,
noting  that tidal forces may vanish at a certain point, unlike in the Schwarzschild spacetime, using Boyer-Lindquist
coordinates. We study in detail the effects of vanishing tidal forces in a body constituted of 
dust infalling along the symmetry axis.
We find the geodesic deviation equations and solve them, numerically and analytically,
for motion along the symmetry axis of the Kerr geometry. We also point out that the intrinsic Gaussian curvature of the event horizon at the symmetry axis is equal to the component along this axis of the tidal tensor
there. 
\end{abstract}

\maketitle

%%%%%%%%%%%%%%%%%%%%%%%%%%%%%%%%%%%%%%%%%%%%%%
\section{Introduction}
\label{Introduction} 
%%%%%%%%%%%%%%%%%%%%%%%%%%%%%%%%%%%%%%%%%
Black Holes (BHs) and Compact Objects (COs) in general are plausible candidates for testing the strong-field regime of General Relativity (GR). The international collaboration Event Horizon Telescope (EHT) released the first image of the shadow of a black hole at the center of the galaxy Messier 87 (M87) \cite{EHT}. The shadow 
of the black hole is consistent with a rotating solution predicted by GR. The first nontrivial exact solution of Einstein's field equations was 
the Schwarzschild geometry, which possesses 
spherical symmetry and describes the spacetime outside a non-rotating and uncharged BH. Another example of a solution with spherical symmetry is the Reissner-Nordstr{\"o}m geometry, which describes the spacetime outside a non-rotating and electrically charged BH. 

Besides spherically symmetric solutions, we have rotating solutions in GR. These solutions possess many interesting features such as the dragging of inertial frames and the existence of stationary limit surfaces. The Kerr spacetime is one example of such rotating solutions, which describes the geometry around electrically uncharged  and rotating BHs. These fully collapsed structures with non-vanishing angular momentum can be associated to many interesting features  in absorption \cite{ACS: 2017, ScalarAbsp-Kerr, ElectrAbsp-Kerr, ABS: 2018, SCT:2019, SCT:2008} and scattering~\cite{SCT:2001, SCT:2008,SCT:InPrep} of fields and superradiant instabilities \cite{SHod:2016,Kerr-Newmaninst:2018}. 

It is well known that a distribution of non-interacting particles in radial free fall towards a Schwarzschild BH would get stretched in the radial direction and compressed in the angular directions \cite{hobson,  dinverno, carroll}. These compression and stretching arise from tidal effects produced by gravitational interaction. In Ref.~\cite{RNTF}
the tidal forces in Reissner-Nordstr{\"o}m spacetime were presented and the geodesic deviation equations 
were solved. There it was noted 
that tidal forces in Reissner-Nordstr{\"o}m spacetime may vanish and change sign  for
certain values of the radial coordinate. Investigations of tidal forces in some other spherically symmetric spacetimes, for instance, regular black holes \cite{RBH}, and Kiselev  black holes \cite{KiselevBH} can also be found in the literature. 

Tidal forces play an important role in astrophysical context, for example, in the effect of tidal disruption of stars, called tidal disruption events (TDE). In 
Ref.~\cite{Luminet and Marck:1984}, Luminet and Marck  investigated how tidal compression acting on the direction orthogonal to the orbital plane induces squeezing on the star when the periastron of the orbit is close to the event horizon of a Schwarzschild BH. Tidal effects for stars orbiting Kerr black holes, at the equatorial plane, have been studied, for instance, in Ref.~\cite{Fishbone} by Fishbone. In Ref.~\cite{I_S_M}, the previous works of Luminet, Marck and Fishbone were extended by formulating the problem in Fermi normal coordinates. Astronomical data for TDE have been presented recently in the literature (see, for instance, Ref.~\cite{TDE}).

In this paper, we investigate the tidal forces in Kerr spacetime. Tidal forces in rotating BH spacetimes have been studied by some authors \cite{Marck:1983,AxisTF,Mahajan:1981,R1,R2,R3,R4,R5,R6}. In particular, the tidal tensor for an arbitrary geodesic was discussed in Ref.~\cite{Marck:1983} using Boyer-Lindquist coordinates. 
Here, we restrict our analysis to the motion along the axis of symmetry of a Kerr BH. 
In this paper we present a new method to obtain the tidal forces along the axis of symmetry of a Kerr black hole. We
use the results of Ref.~\cite{Marck:1983} in Boyer-Lindquist coordinates and take a suitable limit to find the tidal tensor
and geodesic deviation equations, confirming the results of Refs.~\cite{AxisTF,R2,R3,R4}.  It is noteworthy that the tidal tensor 
vanishes for a certain value of the radial coordinate. We investigate the effects of the vanishing tidal forces, numerically 
and analytically,
on the solutions of the geodesic deviation equations as functions of the radial coordinate. 

The timelike geodesic motion along the axis of symmetry is important in astrophysics
 For instance, they may describe the production of very high energy particles and also extra-galactic jets stemming from the center of the galaxies, in which the particles follow a geodesic near the axis of symmetry~\cite{geodesicsaxis}. The relation between jets and tidal forces along the symmetry axis of a Kerr black hole was studied, for instance, in Refs.~\cite{R2,R4,R6}.

The remainder of this paper is structured as follows. In Sec. ~\ref{The Kerr spacetime}, we review the Kerr 
geometry and some of its properties, such as the existence of an ergoregion, and
present the geodesic equations in Kerr spacetime. In Sec.~\ref{Parallel-Propagated tetrads}, we construct the orthonormal and parallel-propagated tetrad for geodesic motion along the rotation axis  by taking a suitable limit of some results of Ref.~\cite{Marck:1983},   
which are summarized in \ref{Marck_tetrad}, and present a new method to find the tidal tensor for the on-axis case. In Sec.~\ref{Gaussian_Curvature}, we note that the tidal forces and the Gaussian curvature of the event horizon 
at the symmetry axis vanish for the same value of the rotation parameter of the Kerr black hole. We analyze this result, and point out that these
quantities are in fact equal along the symmetry axis for any value of the rotation parameter. 
In Sec.~\ref{Solution to the geodesic deviation equation}, we find numerical solutions for the geodesic deviation equation. 
In Sec.~\ref{sec:analytic_solution}, we construct the analytic solutions to the geodesic deviation equations by suitably 
varying the geodesic equations.
In Sec.~\ref{Conclusion}, we present our conclusion. We use the metric signature $(-,+,+,+)$ and set the speed of light and the Newtonian gravitational constant equal to one.

%%%%%%%%%%%%%%%%%%%%%%%%%%%%%%%%%%%%%%%%%%%%%%%%%%%%%%%%%%%%%
\section{Kerr spacetime}
\label{The Kerr spacetime}
%%%%%%%%%%%%%%%%%%%%%%%%%%%%%%%%%%%%%%%%%%%%%%%%%%%%%%%%%%%%%%
The Kerr geometry 
was first presented as an example of algebraically special solution of Einstein's equations, and 
it describes the spacetime around an electrically uncharged and rotating BH \cite{Kerr:1963}. The Kerr line element in Boyer-Lindquist coordinates $(t,r,\theta, \phi)$ is given by \cite{hobson,dinverno,carroll}
\begin{eqnarray}
\nonumber ds^2=&&-\left(1-\frac{2\,M\,r}{\Sigma}\right)dt^2-2\frac{2\,M\,r\,a\,\sin^2\theta}{\Sigma}dt\,d\phi+\frac{\Sigma}{\Delta}dr^2\\
&&+\Sigma\,d\theta^2+\frac{\mathcal{A}}{\Sigma}\sin^2\theta\,d\phi^2, \label{Kerr-metric}
\end{eqnarray}
where $M$ is the mass and $a$ is the angular momentum per unit mass of the Kerr BH. The functions $\Sigma, \Delta$ and $\mathcal{A}$  of the $r$ and $\theta$ coordinates are given by
\begin{eqnarray}
&&\Sigma=r^2+a^2\,\cos^2\theta,\\
&&\Delta=r^2+a^2-2\,M\,r,\\
&&\mathcal{A}=(r^2+a^2)^2-\Delta\,a^2\,\sin^2\theta.
\end{eqnarray}

The Kerr spacetime possesses two horizons. One is the event horizon ($r=r_+$) and the other is a Cauchy horizon 
($r=r_-$), which are solutions of
\begin{equation}
\Delta=r^2+a^2-2\,M\,r=0,
\end{equation}
given explicitly by
\begin{eqnarray}
\label{horinz}&&r_+=M+(M^2-a^2)^\frac{1}{2},\\
\label{cauchy}&&r_-=M-(M^2-a^2)^\frac{1}{2}.
\end{eqnarray}

Note that, in order for Eqs.~\eqref{horinz} and \eqref{cauchy} to be real valued, we must have
$a^2 \leq M^2$.
For 
$a=M$, we have the so-called extreme Kerr BH, for which the Cauchy horizon and the event horizon coincide. Our 
analysis will be restricted by the condition $a^2 \leq M^2$ because otherwise the spacetime would have a naked
singularity.

The curvature singularity in Kerr spacetime can be determined through the evaluation of the Kretschmann scalar, and is located at $\Sigma=0$~\cite{Kerrbook}.
The solution of this equation 
$(r,\theta)=(0,\pi/2)$, is known to be a ring  at the equatorial plane.

Another property of Kerr spacetime is the existence of the so-called ergoregion, which is 
 given by the condition~\cite{Wald}
\begin{equation}
r_+<r<M+(M^2-a^2\,\cos^2\theta)^\frac{1}{2}.
\end{equation}    
Inside the ergoregion an observer cannot remain with fixed $(r, \theta, \phi)$ coordinates, being obliged to co-rotate with the Kerr BH. Along the axis of symmetry of the Kerr BH, the  ergosphere, the outer
limit of the ergoregion,  touches the event horizon.

 As is well known, Carter has found
that Kerr spacetime allows full integrability of the geodesic equations of motion. The equations of timelike geodesic motion in Kerr spacetime are given by \cite{Carter:1968}
\begin{eqnarray}
\label{cartereq0}&&\dot{t}=\frac{(\mathcal{A}\,E-2\,M\,r\,a\,\Phi)}{\Delta\,\Sigma},\\
\label{cartereq1}&&\dot{r}^2=\frac{\left[E(r^2+a^2)-a\,\Phi \right]^2-\Delta(r^2+K)}{\Sigma^2},\\
\label{cartereq2}&&\dot{\theta}^2=\frac{K-a^2\cos^2\theta -\left(a\,E\,\sin\theta-\frac{\Phi}{\sin\theta}\right)^2}{\Sigma^2},\\
\label{cartereq3}&&\dot{\phi}=\frac{\frac{2\,M\,r\,a\,E}{\Sigma}+\left(1-\frac{2\,M\,r}{\Sigma} \right)\frac{\Phi}{\sin^2\theta}}{\Delta}.
\end{eqnarray}
Here, overdots represent differentiation with respect to the proper time. The quantities $E$, $\Phi$ are conserved  along the timelike geodesic related to the energy and angular momentum, respectively, and $K$ is associated to the Carter's constant.
It is clear from Eqs.~\eqref{cartereq2} and \eqref{cartereq3} that $\Phi=0$ for the motion along the rotation axis.

%%%%%%%%%%%%%%%%%%%%%%%%%%%%%%%%%%%%%%%%%%%%%%%%%%%%%%%%%%%
\section{Geodesic equations, parallel-propagated tetrads and tidal tensor along the symmetry axis of a Kerr BH}  
\label{Parallel-Propagated tetrads}
%%%%%%%%%%%%%%%%%%%%%%%%%%%%%%%%%%%%%%%%%%%%%%%%%%%%%%%%%%%%%%%%%%
In various situations, it is convenient to treat problems in GR using a tetrad basis instead of a coordinate basis. Here, we use a tetrad basis in Ref.~\cite{Marck:1983}
to  find 
the tidal tensor and the expression for tidal forces along the symmetry axis of a rotating BH by taking a suitable limit.

 The Kerr geometry possesses several natural observers such as static observers, zero angular momentum observers (ZAMO) and Carter observers (see Ref.~\cite{Semerak:1993} for a detailed review). We use a parallel-propagated tetrad along a geodesic on the axis of symmetry,  using results of Marck~\cite{Marck:1983}, which are summarized in \ref{Marck_tetrad}.
\subsection{Geodesic equations along the symmetry axis of a Kerr BH}
As we stated before, one must have $\Phi=0$ for a geodesic along the symmetry axis with $\theta=0$.  
Then, the requirement $\dot{\theta}=0$ and Eq.~\eqref{cartereq2} imply $K=a^2$.
 With $\Phi=0$, $K=a^2$ and $\theta=0$, Eqs.~\eqref{cartereq0}-\eqref{cartereq3} reduce to
\begin{eqnarray}
\label{carteraximotion0}&&\dot{t}=\frac{\Sigma_0}{\Delta}\,E,\\
\label{carteraximotion1}&&\dot{r}^2
=E^2 - \frac{\Delta}{\Sigma_0},\\
\label{carteraximotion2}&&\dot{\theta}^2=0,\\
\label{carteraximotion3}&&\dot{\phi}=\frac{2\,E\,M\,a\,r}{\Sigma_0\,\Delta}.
\end{eqnarray}
where
\begin{equation}
\label{sigmaaxismotion}\Sigma_0 = \Sigma\mid_{\theta=0}=r^2+a^2.
\end{equation}
We also note
\begin{equation}
\label{aaxismotion}\mathcal{A}\mid_{\theta=0}=(r^2+a^2)^2=\Sigma_0^2.
\end{equation}
Equation~\eqref{carteraximotion3} shows that nearby geodesics are rotating about the symmetry axis with a nonzero
angular velocity relative to the constant-$\phi$ hypersurfaces.

Defining the "Newtonian radial acceleration" $A^R$ as
\begin{equation}
A^{R}\equiv\ddot{r},
\end{equation}
from Eq.~\eqref{carteraximotion1} we find that
\begin{equation}
\label{Newtonianacc}A^{R}=\frac{M\left(a^2-r^2\right)}{(r^2+a^2)^2}.
\end{equation}
Equation \eqref{Newtonianacc} gives the "Newtonian radial acceleration" that the rotating Kerr BH gives
a particle in free fall. 
 In the Schwarzschild limit $a\to 0$ we have
\begin{equation}
\label{Newtonianacca0}A^{R}\mid_{a=0}=-\frac{M}{r^2},
\end{equation}
which coincides with the radial acceleration predicted by Newtonian theory of gravity \cite{Symon}.
We point out that a particle falling freely from rest at $r=b>r_{+}$ would appear to
bounces back at some point $R^{stop}$ 
of the radial coordinate $r$.  
This point is given by determining $E^2$ by substituting $r=b$ and $\dot{r}=0$ into Eq.~\eqref{carteraximotion1} 
and then finding the value of $r$, other than $r=b$, such that $\dot{r}=0$.  Thus we find
\begin{equation}
R^{stop}=\frac{a^2}{b}.
\end{equation}
The point with $r = R^{stop}$ 
is always located inside the Cauchy horizon and, hence, inside the event horizon. 
In fact this particle would emerge in another region of the analytic extension of this spacetime as is well known \cite{MaximalExtensionKerr}.

\subsection{Parallel-propagated tetrads along the symmetry axis of a Kerr BH}

As stated in \ref{Marck_tetrad}, for $\theta=0$  the basis vectors in 
Carter's tetrad $e^\mu_{(a)}$, $a=0,1,2$, given by Eqs.~\eqref{carter0}-\eqref{carter2} 
are in the directions of increasing $t$, $r$ and $\theta$, respectively, whereas $e^\mu_{(3)}$ 
given by Eq.~\eqref{carter3} is in the
direction of decreasing $\phi$.  As a result, it is rotating about the symmetry axis with the angular velocity given by
Eq.~\eqref{carteraximotion3} relative to the constant-$\phi$ hypersurfaces.

The parallel-propagated tetrad $\lambda^\mu_{\hat{\beta}}$, $\beta=0,1,2,3$, can be expanded in 
Carter's tetrad basis as 
\begin{equation}
\label{lambda-mu-b}\lambda^\mu_{\hat{\beta}} = \sum_{a=0}^3 
\lambda_{\hat{\beta}}^{(a)}e^\mu_{(a)},\ \  b=0,1,2,3.
\end{equation}
By letting $K=a^2$ and then 
$\theta=0$ the coefficients $\lambda_{\hat{\beta}}^{(a)}$ presented in \ref{Marck_tetrad} reduce to
\begin{eqnarray}
\nonumber&&\lambda^{(a)}_{\ \hat{0}}=\left(\frac{\Sigma_0}{\Delta}\right)^\frac{1}{2}\,E\,\delta^{(a)}_{(0)}\\
\label{tetrad0}&&-\left(\frac{\Sigma_0}{\Delta}\right)^\frac{1}{2}\left(E^2-\frac{\Delta}{\Sigma_0}\right)^\frac{1}{2}\delta^{(a)}_{(1)}\\
\nonumber&&\lambda^{(a)}_{\ \hat{2}}=-\left(\frac{\Sigma_0}{\Delta}\right)^\frac{1}{2}
\left(E^2-\frac{\Delta}{\Sigma_0}\right)^\frac{1}{2}\delta^{(a)}_{(0)}\\
\label{tetrad1}&&+\left(\frac{\Sigma_0}{\Delta}\right)^\frac{1}{2}\,E\,\delta^{(a)}_{(1)}\\
\label{tetrad2}&&\lambda^{(a)}_{\ \hat{1}}=\tilde{\lambda}^{(a)}_{\ \hat{1}}\,\cos\Psi-\tilde{\lambda}^{(a)}_{\ \hat{3}}\,\sin\Psi,\\
\label{tetrad3}&&\lambda^{(a)}_{\ \hat{3}}=\tilde{\lambda}^{(a)}_{\ \hat{1}}\,\sin\Psi+\tilde{\lambda}^{(a)}_{\ \hat{3}}\cos\Psi,
\end{eqnarray}
where $\delta^{(a)}_{(a')}$ is the Kronecker delta. Here and below, 
we are assuming that the geodesic represents a particle
falling into the black hole and, hence, that $\dot{r} < 0$.
The 4-vectors $\tilde{\lambda}^{(a)}_{\ \hat{1}}$ and $\tilde{\lambda}^{(a)}_{\ \hat{3}}$ are given by
\begin{eqnarray}
&&\tilde{\lambda}^{(a)}_{\ \hat{1}}=E\,\delta^{(a)}_{(2)}\mp\sqrt{1-E^2}\,\delta^{(a)}_{(3)},\\
&&\tilde{\lambda}^{(a)}_{\ \hat{3}}=\pm \sqrt{1-E^2}\,\delta^{(a)}_{(2)}+E\delta^{(a)}_{(3)},
\end{eqnarray}
where the upper (lower) sign corresponds to $\dot{\theta}$ being positive (negative).

For geodesic motion along the symmetry axis for which $\Phi=0$ and $K=a^2$ the proper-time derivative of
$\Psi$ given by Eq.~\eqref{psi} vanishes.
This implies that the rotation angle $\Psi$ is a constant with respect to the proper time. 
This in turn implies we may take 
$\lambda^{(a)}_{\ \hat{1}} = \cos\beta\,\delta^{(a)}_{(2)}+\sin\beta\,\delta^{(a)}_{(3)}$ and 
$\lambda^{(a)}_{\ \hat{3}} = - \sin\beta\,\delta^{(a)}_{(2)} + \cos\beta\,\delta^{(a)}_{(3)}$ with arbitrary 
constant angle $\beta$.  Hence, the parallel-transported 
tetrad is not rotating relative to the Carter tetrad and, therefore, is 
rotating about the symmetry axis with the angular velocity given by
Eq.~\eqref{carteraximotion3} relative to the constant-$\phi$ hypersurfaces.
Thus, $\lambda^\mu_{\hat{1}}$ and $\lambda^\mu_{\hat{2}}$ are any orthonormal 
spatial vectors 
perpendicular to the rotation axis that are rotating with this angular velocity.\footnote{If one takes the limit 
$\theta\to 0$ first and then the limit $K\to a^2$, the angular velocity $\dot{\Psi}$ of 
$(\lambda_{\hat{1}}^\mu,\lambda_{\hat{3}}^\mu)$ relative to 
$(\tilde{\lambda}_{\hat{1}}^\mu,\tilde{\lambda}_{\hat{3}}^\mu)$ will be nonzero.  However, in this case the angular velocity
of the latter relative to the Carter tetrad will be nonzero as well, and these two angular velocities cancel out.  Hence,
the parallel-transported tetrad $\lambda_{\hat{\beta}}^{(a)}$ is rotating with the Carter tetrad with angular velocity 
$\dot{\phi}$ given by Eq.~\eqref{carteraximotion3} relative to the constant-$\phi$ hypersurfaces irrespective of the
order of limits $K\to a^2$ and $\theta\to 0$ as it should.}

Tsoubelis and Economou~\cite{AxisTF} calculated the rotation of the parallel-transported tetrad along the symmetry
axis in Kerr-Schild coordinates.
The angular velocity of the parallel-transported tetrad along the symmetry axis in these coordinates is
\begin{equation}
\dot{\phi}_{KS} = \frac{2Mar}{\Sigma_0\Delta}\left( E - \sqrt{E^2 - 1 - \frac{\Delta}{\Sigma_0}}\right),
\label{differente}
\end{equation}
for $\dot{r} < 0$, 
where $\dot{\phi}_{KS}$ is defined from the Kerr-Schild coordinates $(x,y)$ by $x+iy = \rho\,e^{i\phi_{KS}}$, for
$\dot{r} < 0$.  This result can be shown to agree with Eq.~\eqref{carteraximotion3} by noting that
\begin{equation}
\dot{\phi}_{KS} = \dot{\phi} + \frac{2Mar}{\Sigma_0\Delta}\dot{r},
\end{equation}
which follows from the relation
\begin{equation}
x+iy = (r+ia)\sin\theta \exp\left( i\phi + i\int \frac{a}{\Delta}dr\right).
\end{equation}
(See, e.g.\ Ref.~\cite{Townsend}.) 

\subsection{Geodesic deviation equations and the tidal tensor along the symmetry axis of a Kerr BH}
\label{Sec3.3}
We now write down 
the geodesic deviation equations for a particle moving along the axis of symmetry of the Kerr spacetime. For
a spacelike vector $\eta^\mu = \sum_{\alpha=1}^3 \eta^{\hat{\alpha}}\lambda^\mu_{\hat{\alpha}}$, where
$\lambda^\mu_{\hat{\alpha}}$ are defined by Eq.~\eqref{lambda-mu-b},
that represents the distance between two infinitesimally close particles following geodesics, the equation for the relative 
acceleration is \cite{dinverno}
\begin{equation}
\label{tidalacce}\frac{D^2\eta^{\hat{\alpha}}}{D\,\tau^2}=-K^{\hat{\alpha}}_{\ \hat{\beta}}\,\eta^{\hat{\beta}},
\end{equation}
where $K^{\hat{\alpha}}_{\ \hat{\beta}}$ is the tidal tensor, defined in terms of the curvature tensor, as in 
Eq.~\eqref{tidaltensorincartercomp}. Here $\hat{\alpha},\hat{\beta}= 1,2,3$.

Using Eqs. \eqref{tetrad0}-\eqref{tetrad3} and the non-zero components of the Riemann curvature tensor, we find the non-vanishing components of the tidal tensor to be
\begin{eqnarray}
\label{tidaltensor1}K_{\hat{1}\,\hat{1}}=\frac{M\,r}{(r^2+a^2)^3}(r^2-3\,a^2),\\
\label{tidaltensor2}K_{\hat{2}\,\hat{2}}=\frac{-2\,M\,r}{(r^2+a^2)^3}(r^2-3\,a^2),\\
\label{tidaltensor3}K_{\hat{3}\,\hat{3}}=\frac{M\,r}{(r^2+a^2)^3}(r^2-3\,a^2),
\end{eqnarray}
 which agrees with the results found in Ref.~\cite{AxisTF} using Kerr-Schild coordinates. We emphasize that the off-diagonal terms of the tidal tensor vanish 
for geodesic motion along the BH rotation axis. We can also obtain the results \eqref{tidaltensor1}-\eqref{tidaltensor3} straightforwardly from Eqs.~\eqref{Tidaltensor11}-\eqref{I2tidalforce}. 
 The geodesic deviation equations are readily obtained 
by substituting Eqs.~\eqref{tidaltensor1}-\eqref{tidaltensor3} into Eq.~\eqref{tidalacce} to be
\begin{eqnarray}
\label{radialtf}&&\frac{D^2\eta^{\hat{1}}}{D\tau^2}=-\frac{M\,r}{(r^2+a^2)^3}(r^2-3a^2)\,\eta^{\hat{1}},\\
\label{angulartf1}&&\frac{D^2\eta^{\hat{2}}}{D\tau^2}=\frac{2\,M\,r}{(r^2+a^2)^3}(r^2-3a^2)\,\eta^{\hat{2}},\\
\label{angulartf2}&&\frac{D^2\eta^{\hat{3}}}{D\tau^2}=-\frac{M\,r}{(r^2+a^2)^3}(r^2-3a^2)\,\eta^{\hat{3}}.
\end{eqnarray}

From Eqs.~\eqref{radialtf}-\eqref{angulartf2}, we see that the expressions for tidal forces in Kerr spacetime, for a motion along the BH rotation axis, are the same in  $\eta^{\hat1}$ and  $\eta^{\hat3}$ directions,
i.e.\ the directions perpendicular to the rotation axis.  
In the limit $a\rightarrow 0$ we recover the well-known result for 
Schwarzschild tidal forces for radially infalling observers \cite{dinverno}:
\begin{eqnarray}
\lim_{a\rightarrow 0}&&\frac{D^2\eta^{\hat{1}}}{D\tau^2}=-\frac{M}{r^3},\\
\lim_{a\rightarrow 0}&&\frac{D^2\eta^{\hat{2}}}{D\tau^2}=\frac{2\,M}{r^3},\\
\lim_{a\rightarrow 0}&&\frac{D^2\eta^{\hat{3}}}{D\tau^2}=-\frac{M}{r^3},
\end{eqnarray}
noting that $\eta^{\hat{2}}$ represents the geodesic deviation in the direction of motion and
$\eta^{\hat{1}}$ and $\eta^{\hat{3}}$ represent that in the perpendicular directions.
 It is interesting that the tidal forces in all
directions vanish at the same value of radial coordinate, namely $R_{0}\equiv \sqrt{3}\,a$.  
For values of the rotation parameter in the range $\frac{\sqrt{3}}{2}\,M\leq a\leq M$, the point with 
$r=R_{0}$ is located outside the event horizon.   There is evidence of rapidly spinning black holes with rotation parameter greater than 
$a=\frac{\sqrt{3}}{2}\,M\approx 0.87\,M$. 
Therefore, the vanishing tidal forces can be observed in nature, in principle. (See, for example, 
Refs.~\cite{rspinningbh,rspinningbh2}, in which observational data of a BH are shown to
imply the rotation parameter to be approximately $a\approx 0.92\,M$.) 

\begin{center}
\begin{figure}[h!]
\center
\includegraphics[scale=0.8]{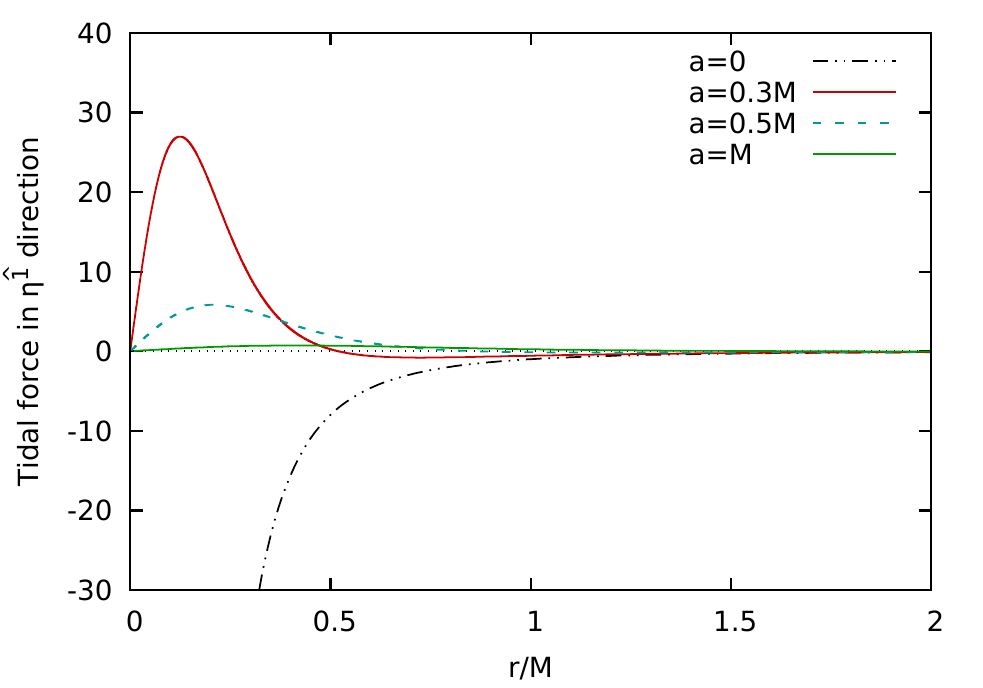}
\caption{Tidal force for motion along the symmetry axis in Kerr spacetime
in the direction perpendicular to the axis, i.e.\ in the direction of $\eta^{\hat{1}}$ and $\eta^{\hat{3}}$, given
by \eqref{radialtf} and \eqref{angulartf2}
with different choices of $a$.}
\label{rtf}
\end{figure}
\end{center}

\begin{center}
\begin{figure}[h!]
\center
\includegraphics[scale=0.8]{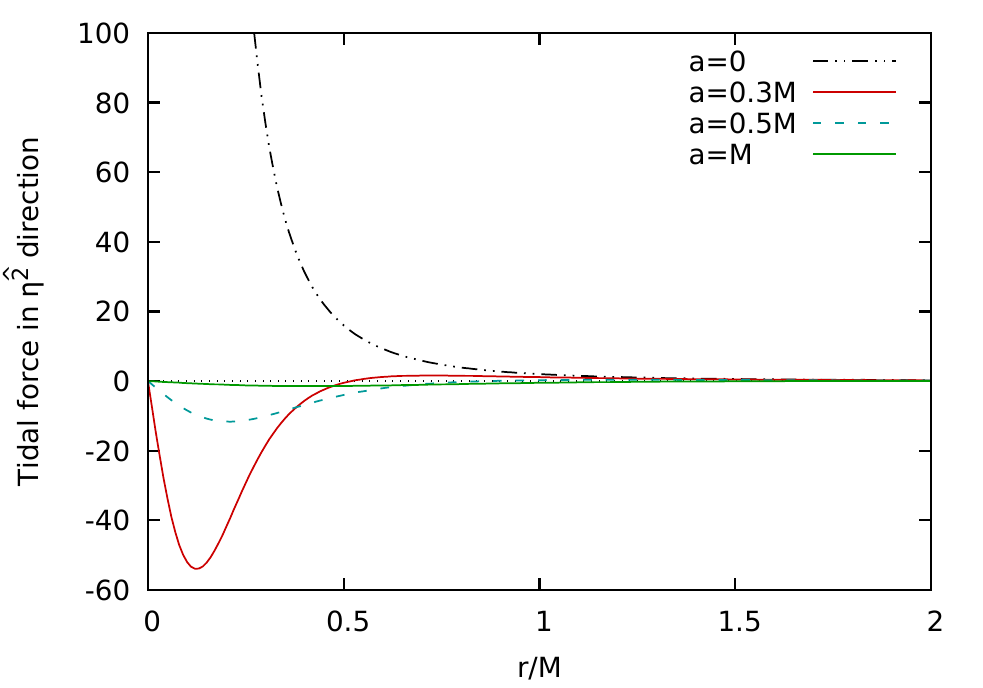}
\caption{Tidal force in the $\eta^{\hat{2}}$ direction, for motion along the symmetry axis in Kerr spacetime, given by Eq. \eqref{angulartf1}, with different choices of $a$.}
\label{tff}
\end{figure}
\end{center}

\begin{center}
\begin{figure}[h!]
\center
\includegraphics[scale=0.9]{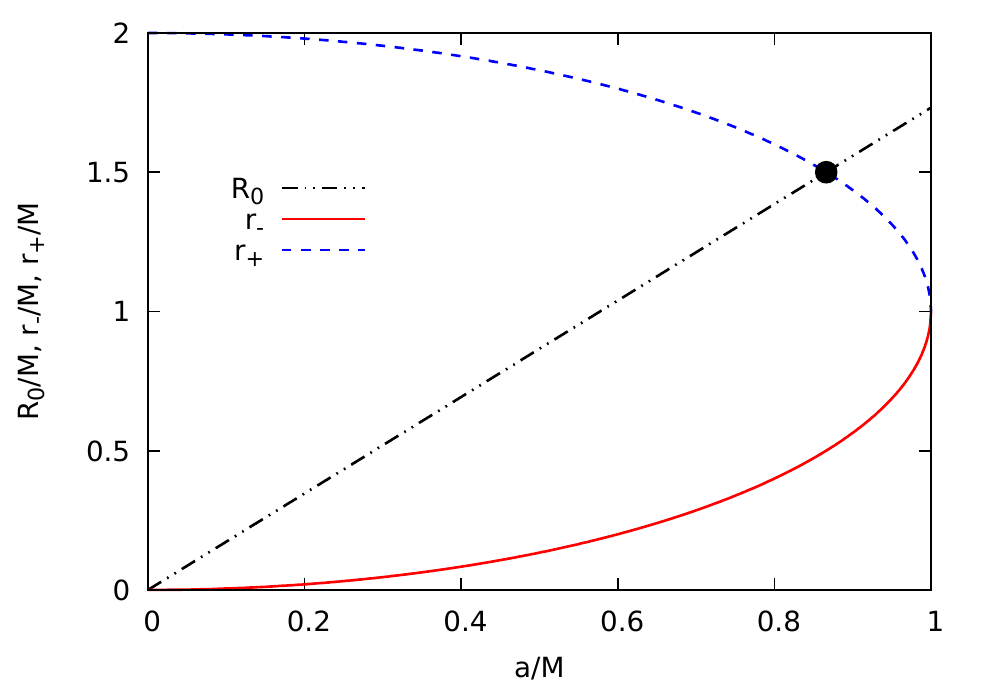}
\caption{$R_0$, $r_-$ and $r_+$ plotted as functions of $a$. The black dot in the plot represents the intersection between $R_0$ and $r_+$, which is
at $a/M=\sqrt{3}/2$, where the tidal forces invert their sign at the event horizon. For values of $a/M$ greater than 
$\sqrt{3}/2$, the tidal forces invert their sign outside the event horizon.}
\label{tidal_properties}
\end{figure}
\end{center}

  In Figs.~\ref{rtf} and \ref{tff}, we show the behavior of the tidal forces in the 
Kerr geometry for observers moving along the BH rotation axis, given by Eqs.~\eqref{radialtf}-\eqref{angulartf2}. In Fig.~\ref{tidal_properties} we plot $R_{0}$, $r_-$ and $r_+$ as a function of the BH rotation parameter.
\hspace{0.3cm} The tidal forces diverge as the body approaches the singularity in the case $a=0$, which is a well-known behavior in Schwarzschild case.  Note that for the extreme Kerr BH, the magnitude of the tidal forces is far smaller than the other cases.
%%%%%%%%%%%%%%%%%%%%%%%%%%%%%%%%%%%%%%%%%%%%%%%%%%%%%%%%%%%%%%%%%%%%%%%%%%%%%%%%%%%%%%%%%%%%%%%%%%%%%%%%%%%%%%%%%%
\section{Tidal forces and Gaussian curvature at the event horizon}
\label{Gaussian_Curvature}
The Gaussian curvature of the event horizon of a Kerr black hole was first studied in Ref.~\cite{Smarr}. It was shown that for $a=\sqrt{3}/2\,M$, the Gaussian curvature of the event horizon becomes zero along the axis of symmetry, while for $a>\sqrt{3}/2\,M$  the Gaussian curvature is negative and the isometric embedding of the event horizon surface in a three dimensional Euclidean space ($\mathbb{E}^3$) is not possible.

As pointed out in Sec.~\ref{Sec3.3}, the tidal forces vanish at the event horizon for $a/M=\sqrt{3}/2$. This is exactly the same value of $a$ where the Gaussian curvature of the event horizon vanishes along the axis of symmetry. This is not a coincidence, since the Gaussian curvature is equal to the tidal force in $\eta^{\hat{2}}$ direction at the event horizon.

\begin{figure*}
  \centering
  \subfigure[]{\includegraphics[scale=0.5]{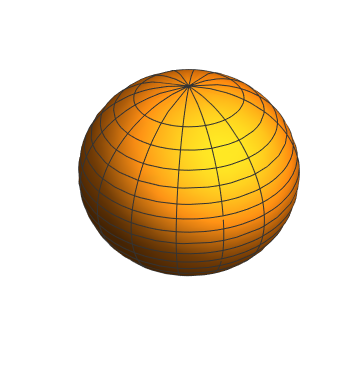}\label{a0}}\quad
  \subfigure[]{\includegraphics[scale=0.5]{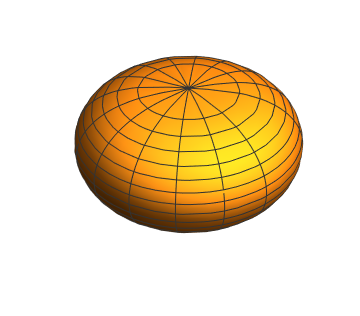}\label{b0}}
\subfigure[]
{\includegraphics[scale=0.5]{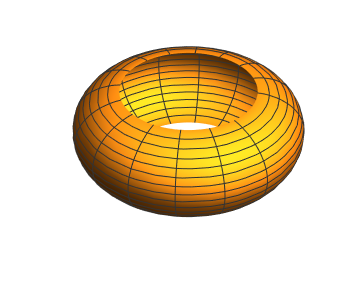}\label{c0}}
\subfigure[]
{\includegraphics[scale=0.5]{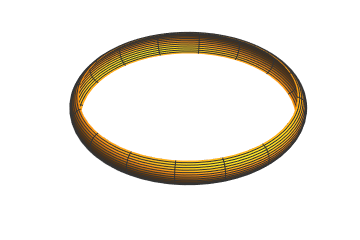}\label{d0}}
\caption{Embedding of the event horizon in $\mathbb{E}^3$ for different values of $a$. 
The rotation parameter has been chosen 
as follows: \text{(a) } $a=0.5\,M , \text{(b) } \sqrt{3}/2\,M, 
\text{(c) }0.92\,M,\text{ and } \text{(d) } M$. In Fig. 4 (b), we note that the Gaussian curvature and the tidal forces are zero at the event horizon along the axis of symmetry. We followed Ref.~\cite{Smarr} in order to construct the embedding presented in this figure.  }\label{Fig.5}
\end{figure*}

The 2-surface with $t=\text{const}$ and $r=r_+$ is described by
\begin{eqnarray}
\label{S2} d\sigma^2=&&\left(r_+^2+a^2\,\cos^2\theta\right)\,d\theta^2
+\frac{\sin^2\theta\,\left(r_+^2+a^2\right)^2}{\left(r_+^2+a^2\,\cos^2\theta\right)}\,d\phi^2.
\end{eqnarray}
By means of the Gauss's Theorema Egregium, the Gaussian curvature K is given by \cite{Dif_geo}
\begin{eqnarray}
\label{Gauss_theo}K=\frac{R^{(\sigma)}_{\ \theta\,\phi\,\theta\,\phi}}{\sigma},
\end{eqnarray} 
where $R^{(\sigma)}_{\ \theta\,\phi\,\theta\,\phi}$ is the only independent component of the Riemann tensor, and $\sigma$ is the determinant of the metric tensor on the 2-surface described by Eq.~\eqref{S2}. Using Eq.~\eqref{Gauss_theo}, we find that
\begin{equation}
\label{Gaussian_Curv}K(\theta)=\frac{2\,M\,r_+\,\left(r_+^2-3\,a^2\,\cos^2\theta\right)}{\left(r_+^2+a^2\,\cos^2\theta\right)^3},
\end{equation} 
which is equal to the tidal force in the $\eta^{\hat{2}}$ direction along the axis of symmetry [see Eq.~\eqref{angulartf1}]. From Eq.~\eqref{Gaussian_Curv}, we see that $K(0)=0$ for $a/M=\sqrt{3}/2$.
%\begin{eqnarray}
%K(0)=0,
%\end{eqnarray}
This explains the fact that the tidal forces and the Gaussian curvature are zero at the event horizon, along the axis of symmetry. For $a/M> \sqrt{3}/2 $, the tidal forces along this axis, as well as the Gaussian curvature, will be negative at the event horizon. The embedding of the event horizon surface in $\mathbb{E}^3$ is shown in Fig.~\ref{Fig.5}, for different values of $a$. For $a/M> \sqrt{3}/2 $ we note the appearance of regions around the rotation axis, which cannot be embedded in the three-dimensional Euclidean space. The tidal forces will be studied
in detail for the case $a/M> \sqrt{3}/2$ 
in Secs.~\ref{Solution to the geodesic deviation equation} and \ref{sec:analytic_solution}. 

%%%%%%%%%%%%%%%%%%%%%%%%%%%%%%%%%%%%%%%%%%%%%%%%%%%%%%%%%%%%%%%%%%%%%%%%%%%%%%%%%%%%%%%%%%%%%%%%%%%%%%%%%%%%%%%%%%
\section{Numerical solutions of the geodesic deviation equations in Kerr spacetime for motion along the symmetry axis}
\label{Solution to the geodesic deviation equation}
%%%%%%%%%%%%%%%%%%%%%%%%%%%%%%%%%%%%%%%%%%%%%%%%%%%%%%%%%%%%%%%%%

Now we solve the geodesic deviation equations presented in Sec. \ref{Parallel-Propagated tetrads} numerically
in order to find the deviation 
vector $\eta^{\hat{\alpha}}$ in terms of the radial coordinate $r$. From Eqs.~\eqref{radialtf}-\eqref{angulartf2}, we obtain a differential equation with respect to the radial coordinate (as independent variable), using that
\begin{equation}
\label{chainrule}\frac{d^2\eta^{\hat{a}}}{d\tau^2}=\frac{dr}{d\tau}\frac{d}{dr}\left(\dot{r}\frac{d\eta^{\hat{a}}}{d\,r}\right)=\dot{r}^2\frac{d^2\eta^{\hat{a}}}{dr^2}+\frac{1}{2}\frac{d \dot{r}^2}{d r}\, \frac{d\eta^{\hat{a}}}{dr}.
\end{equation}
Substituting the expression~\eqref{carteraximotion1} for $\dot{r}^2$ 
and its derivative with respect to the radial coordinate into 
Eq.~\eqref{chainrule}, we obtain 
\begin{equation}
\label{chainrule2}\frac{d^2\eta^{\hat{a}}}{d\tau^2}=\left(E^2-1+\frac{2\,M\,r}{r^2+a^2}\right)\frac{d^2\eta^{\hat{a}}}{d\,r^2}+\frac{M(a^2-r^2)}{(r^2+a^2)^2}\frac{d\eta^{\hat{a}}}{dr}.
\end{equation}
By substituting Eq. \eqref{chainrule2} into Eqs.~\eqref{radialtf}-\eqref{angulartf2}, we find
\begin{eqnarray}
\nonumber&&\left(E^2-1+\frac{2\,M\,r}{r^2+a^2}\right)\frac{d^2\eta^{\hat{i}}}{d\,r^2}+\frac{M(a^2-r^2)}{(r^2+a^2)^2}\frac{d\eta^{\hat{i}}}{dr}\\
\label{eqdifr}&&+\frac{M\,r}{(r^2+a^2)^3}(r^2-3a^2)\,\eta^{\hat{i}}=0,\\
\nonumber&&\left(E^2-1+\frac{2\,M\,r}{r^2+a^2}\right)\frac{d^2\eta^{\hat{2}}}{d\,r^2}+\frac{M(a^2-r^2)}{(r^2+a^2)^2}\frac{d\eta^{\hat{2}}}{dr}
\label{eqdift}\\&&
-\frac{2\,M\,r}{(r^2+a^2)^3}(r^2-3a^2)\,\eta^{\hat{2}}=0,
\end{eqnarray}
 where $i=1,3$.
 
 We solve numerically the ordinary differential equations \eqref{eqdifr} and \eqref{eqdift}. 
We consider two types of initial conditions representing dust of particles starting at $r=b > r_+$ with the 
center being at rest. The first type of initial conditions (IC1) is given by
  \begin{eqnarray}
\label{IC11}  &&\eta^{\hat{\alpha}}(b)= 1,\\
\label{IC12}   &&\left. \frac{d\eta^{\hat{\alpha}}}{d\tau}\right|_b=0,
  \end{eqnarray}
which is associated to a body constituted of dust released with no internal motion at $r=b$. The second type of initial condition (IC2) is given by
  \begin{eqnarray}
\label{IC21}  &&\eta^{\hat{\alpha}}(b)= 0,\\
\label{IC22}   &&\left. \frac{d\eta^{\hat{\alpha}}}{d\tau}\right|_b= 1,
  \end{eqnarray}
  which is associated to a body constituted of dust ``exploding'' from a point at $r=b$  on the symmetry axis.
In the next subsections we treat the solutions of Eqs.~\eqref{eqdifr} and \eqref{eqdift} in detail. 
\subsection{The $\eta^{\hat{i}}$ components of the deviation vector}  
\label{eta13}
\begin{center}
\begin{figure}[h!]
\center
\includegraphics[scale=0.65]{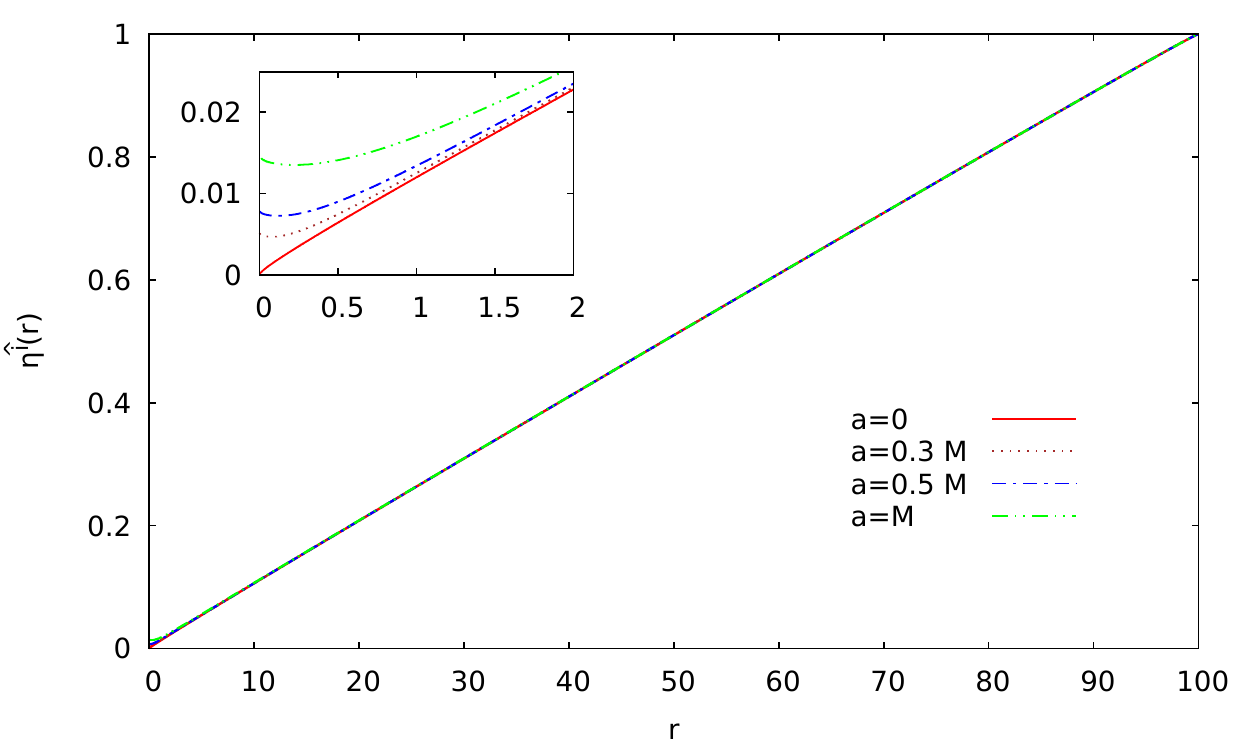}
\caption{The $\eta^{\hat{i}}$ components of the deviation vector for different choices of $a$. In this figure, we have chosen $b=100M$ with initial conditions IC1 described in Eqs.~\eqref{IC11} and  \eqref{IC12}.}
\label{eta1IC1bcons_fig}
\end{figure}
\end{center}

\begin{center}
\begin{figure}[h!]
\center
\includegraphics[scale=0.65]{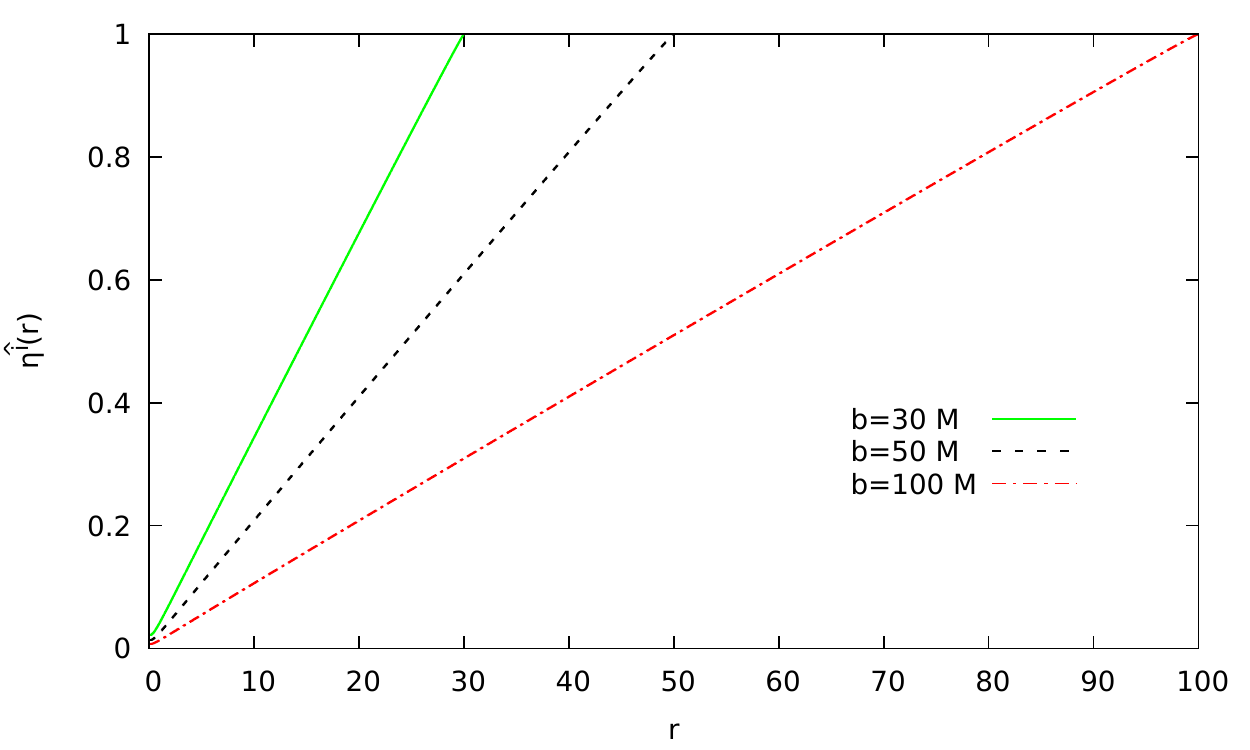}
\caption{The $\eta^{\hat{i}}$ components of the deviation vector for different choices of $b$. In this figure we have chosen $a=0.5\,M$ with initial conditions IC1 described in Eqs.~\eqref{IC11} and  \eqref{IC12}.}
\label{eta1IC1acons_fig}
\end{figure}
\end{center}

For the components perpendicular to the symmetry axis, $\eta^{\hat{i}}$, the solutions of Eq. \eqref{eqdifr}  with the IC1 are plotted in 
Figs.~\ref{eta1IC1bcons_fig} and \ref{eta1IC1acons_fig}.  

In Fig.~\ref{eta1IC1bcons_fig}, we let $b=100\,M$, $\eta^{\hat{i}}(b)=1$ and $M=1$, and plot the $\eta^{\hat{i}}$ component of the deviation vectors for various choices of the rotation parameter $a$ of the Kerr BH. The behavior  of $\eta^{\hat{i}}$ is essentially the same for different values of $a$ at large $r$, as can be seen in Fig.~\ref{eta1IC1bcons_fig}. 
This is because all particles released from rest fall toward the BH for large $r$.
Besides that, we see that during the infall from $r=b$ to the event horizon at $r=r_+$, the component 
$\eta^{\hat{i}}$ always decreases. Also, in Fig.~\ref{eta1IC1bcons_fig} we see that for $a=0$ the body
shrinks to zero size at the origin of the radial coordinate, while for $a \neq 0$ the body maintains a finite size at $r=R^{stop}$. This behavior reflects the tidal force changing from compressing to stretching at $r=\sqrt{3}\,a$.
In Fig.~\ref{eta1IC1acons_fig}, we let 
$a=0.5M$, $\eta^{\hat{i}}(b)=1$ and $M=1$, and plot the $\eta^{\hat{i}}$ component of the 
deviation vectors
for various choices of $b$. The behavior  of $\eta^{\hat{i}}$ is essentially the same as in Fig.~\ref{eta1IC1bcons_fig}, and  as we decrease the value of $b$, $\eta^{\hat{i}}$ decreases faster.
This is, again, because all particles released from rest fall toward the BH for large $r$.

\begin{center}
\begin{figure}[h!]
\center
\includegraphics[scale=0.66]{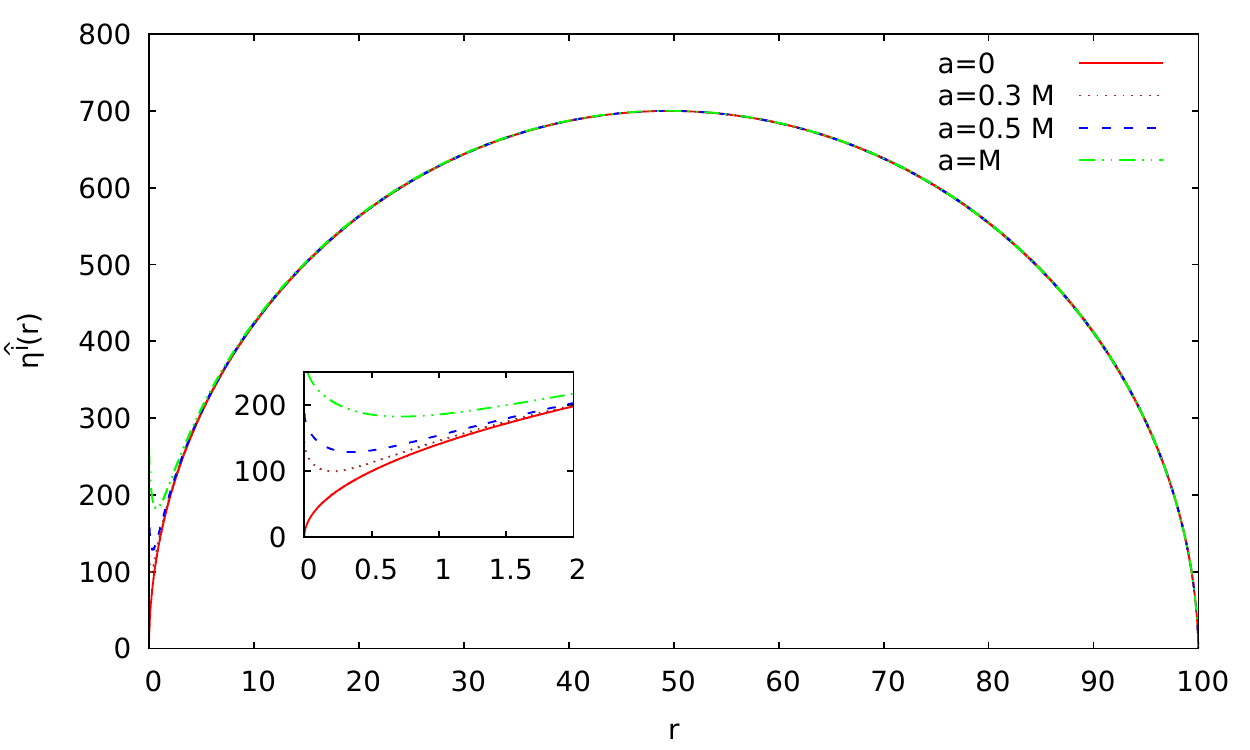}
\caption{The $\eta^{\hat{i}}$ components of the deviation vector for different choices of $a$. In this figure, we have chosen $b=100M$, with initial conditions IC2 described in Eqs.~\eqref{IC21} and  \eqref{IC22}.}
\label{eta1IC2bcons_fig}
\end{figure}
\end{center}

\begin{center}
\begin{figure}[h!]
\center
\includegraphics[scale=0.66]{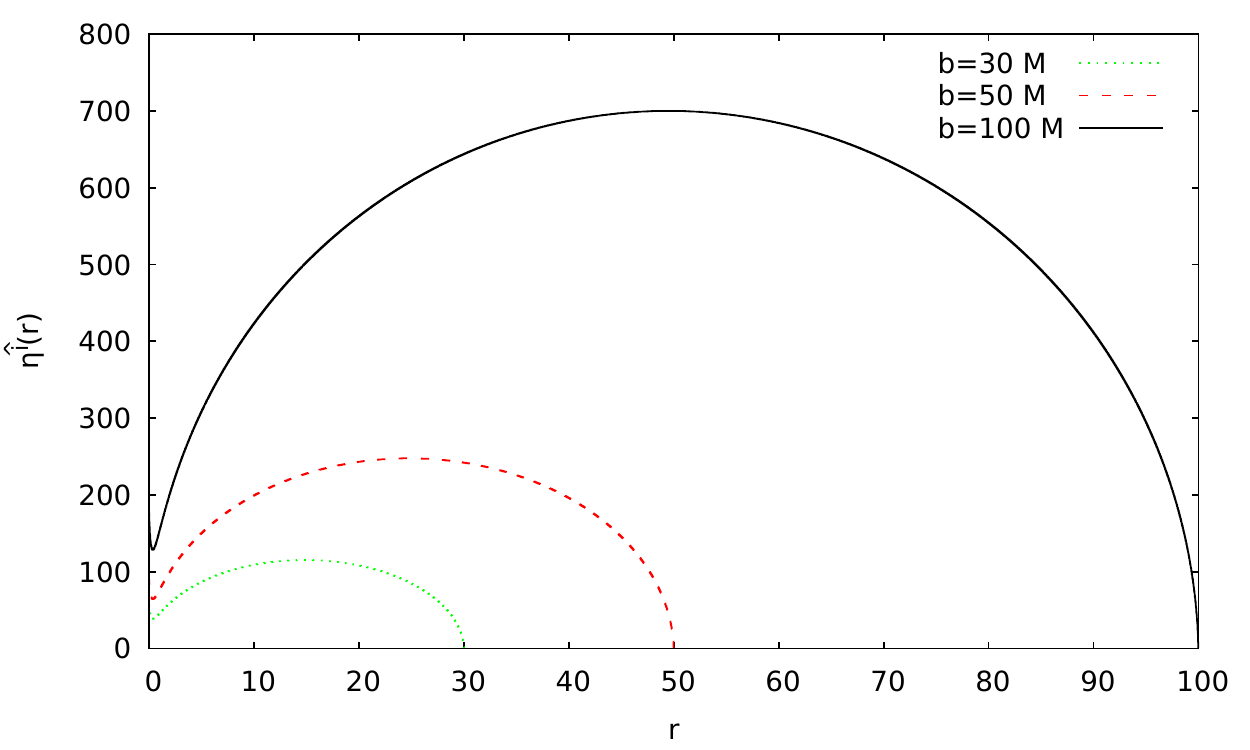}
\caption{The $\eta^{\hat{i}}$ components of the deviation vector for different choices of $b$. In this figure, we have chosen $a=0.5\,M$, with initial conditions IC2, described in Eqs.~\eqref{IC21} and  \eqref{IC22}.}
\label{eta1IC2acons_fig}
\end{figure}
\end{center}

For the component $\eta^{\hat{i}}$ the solutions with the IC2 are plotted in 
Figs.~\ref{eta1IC2bcons_fig} and \ref{eta1IC2acons_fig}.

In Fig.~\ref{eta1IC2bcons_fig}, we  let $b=100M$, $\frac{d\eta^{\hat{i}}}{d\tau}\mid_{r=b}=1$, 
$M=1$, and plot the $\eta^{\hat{i}}$ component of the deviation vectors 
 for various choices of the rotation parameter $a$ of the Kerr black hole. The behavior  of $\eta^{\hat{i}}$ is essentially the same for different values of $a$ at large $r$. We see that during the infall from $r=b$ to the event horizon at $r=r_+$, the component 
$\eta^{\hat{i}}$ initially increases, reaches a maximum around $r=b/2$ and then decreases until $r$ is
very close to $R^{stop}$. For the Schwarzschild case ($a=0$), this behavior is associated to the compressing tidal forces in angular directions.
For $a=0$, as the body approaches the origin of the radial coordinate, the body shrinks to zero size 
 while for $a\neq 0$ it keeps a finite size at $r=R^{stop}$. 
It can be seen that the deviation vector increases near $r=R^{stop}$ because of the change of the tidal force
from compressing to stretching at $r=\sqrt{3}\,a$.
In Fig.~\ref{eta1IC2acons_fig} we let $a=0.5M$, $\frac{d\eta^{\hat{i}}}{d\tau}\mid_{r=b}=1$
 and $M=1$, and plot the $\eta^{\hat{i}}$ component of the deviation vectors for various choices of $b$. The behavior of $\eta^{\hat{i}}$ is essentially the same as in Fig.~\ref{eta1IC2bcons_fig}, and as we decrease the value of $b$, the maximum value of $\eta^{\hat{i}}$ decreases.

%%%%%%%%%%%%%%%%%%%%%%%%%%%%%%%%%%%%%%%%%%%%%%%%%%%%%%%%%%%%%%%%%%%%%%%%%%%%%%%%%%%%%%%%%%%%%%%%%%%%%%%%%%%%%%%%%%
\subsection{The $\eta^{\hat{2}}$ component of the deviation vector}
\begin{center}
\begin{figure}[h!]
\center
\includegraphics[scale=0.65]{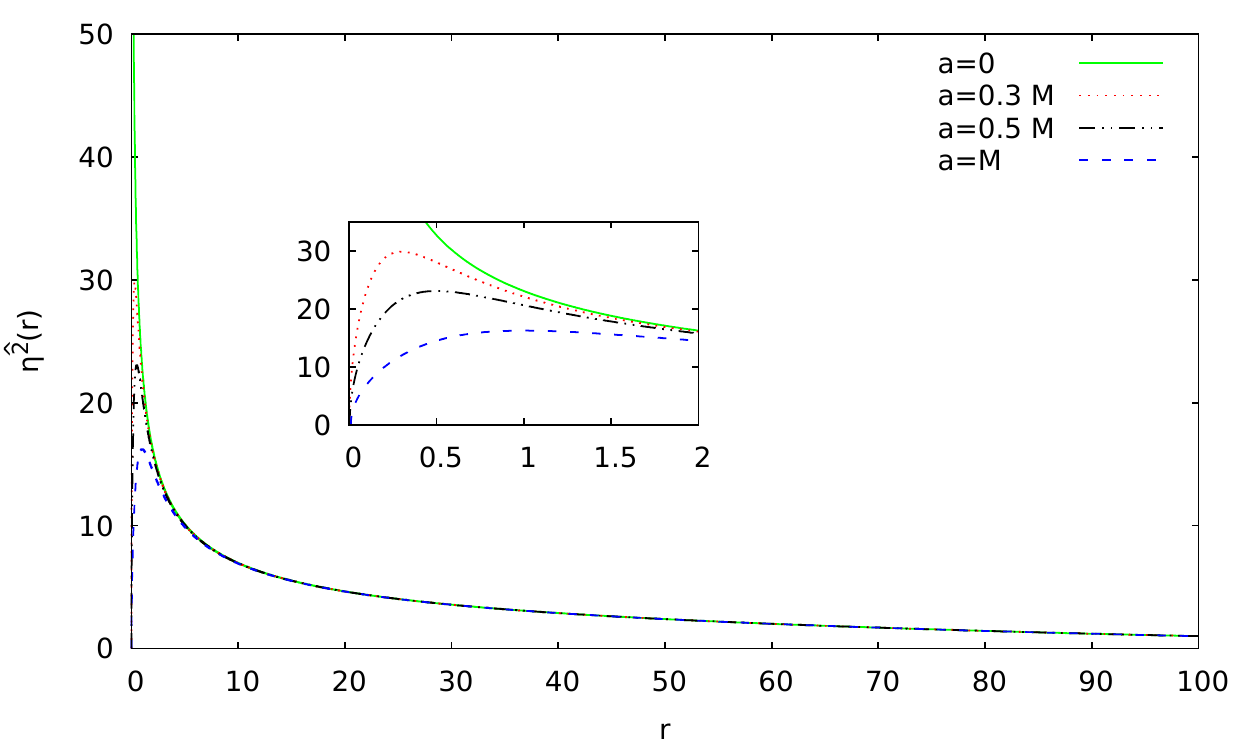}
\caption{The $\eta^{\hat{2}}$ component of the deviation vector, for different choices of $a$. In this figure, we have chosen $b=100M$, with initial conditions IC1, described in Eqs.~\eqref{IC11} and \eqref{IC12}. }
\label{eta2IC1bcons}
\end{figure}
\end{center}

\begin{center}
\begin{figure}[h!]
\center
\includegraphics[scale=0.65]{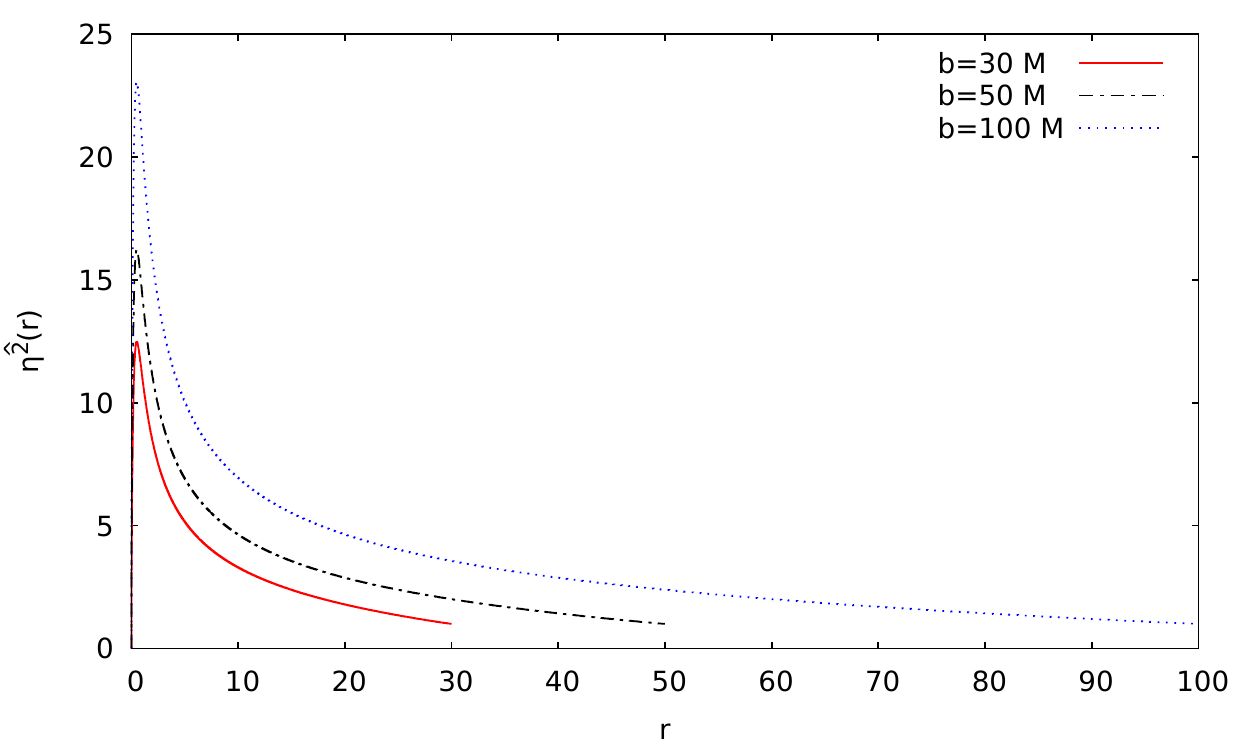}
\caption{The $\eta^{\hat{2}}$ component of the deviation vector, for different choices of $b$. In this figure, we have chosen $a=0.5\,M$, with initial conditions IC1, described in Eqs.~\eqref{IC11} and \eqref{IC12}. }
\label{eta2IC1acons}
\end{figure}
\end{center}

%%%%%%%%%%%%%%%%%%%%%%%%%%%%%%%%%%%%%%%%%%%%%%%%%%%%%%%%%%%%%%%%%%%%%%%%%%%%%%%%%%%%%%%
For the radial component $\eta^{\hat{2}}$ the solutions with the IC1 are plotted in 
Figs.~\ref{eta2IC1bcons} and \ref{eta2IC1acons}.


In Fig.~\ref{eta2IC1bcons}, we let $b=100\,M$, $\eta^{\hat{2}}(b)=1$, $M=1$ and plot the 
$\eta^{\hat{2}}$ component of the deviation vector  for various choices of the rotation parameter $a$ of the Kerr black hole. It can be seen from 
Fig.~\ref{eta2IC1bcons} that the behavior of $\eta^{\hat{2}}$ is essentially the same for different values of $a$ at large $r$. Besides that, for the Schwarzschild case ($a=0$) the component $\eta^{\hat{2}}$ 
tends to infinity as the body approaches the origin of the radial coordinate, which is the location of the singularity. For $a\neq 0 $, the component $\eta^{\hat{2}}$ initially increase, reaches a maximum value and start decreasing
near the BH and becomes very small at $r=R^{stop}$. This is due to the tidal force changing from stretching to
compressing at $r=\sqrt{3}\,a$. 
In Fig.~\ref{eta2IC1acons} we let $a=0.5\,M$, $\eta^{\hat{2}}(b)=1$, $M=1$ and plot 
the $\eta^{\hat{2}}$ component of the deviation vector  for various choices of $b$. The behavior of $\eta^{\hat{2}}$ is essentially the same as in Fig.~\ref{eta2IC1bcons}, and as we decrease the value of $b$, the maximum value of $\eta^{\hat{2}}$ also decreases.

\begin{center}
\begin{figure}[h!]
\center
\includegraphics[scale=0.66]{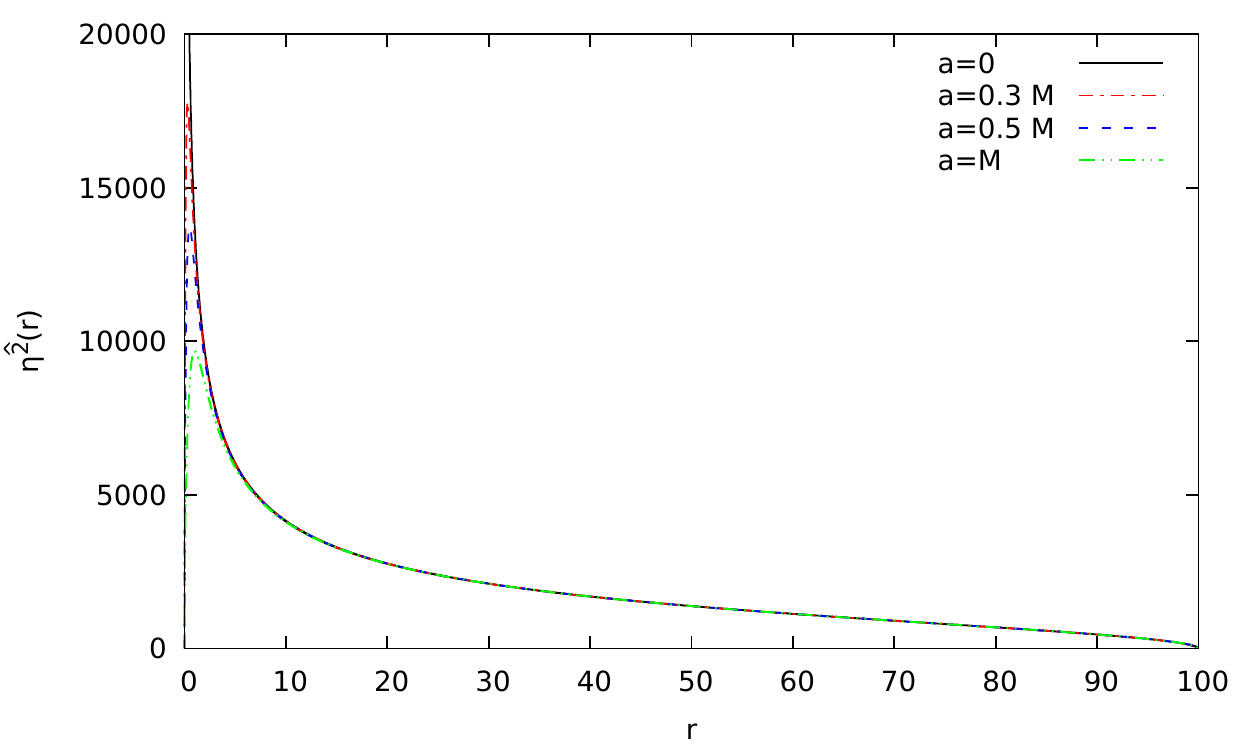}
\caption{The $\eta^{\hat{2}}$ component of the deviation vector, for different choices of $a$. In this figure, we have chosen $b=100M$, with initial conditions IC2, described in Eqs.~\eqref{IC21} and \eqref{IC22}. }
\label{eta2IC2bcons}
\end{figure}
\end{center}

\begin{center}
\begin{figure}[h!]
\center
\includegraphics[scale=0.66]{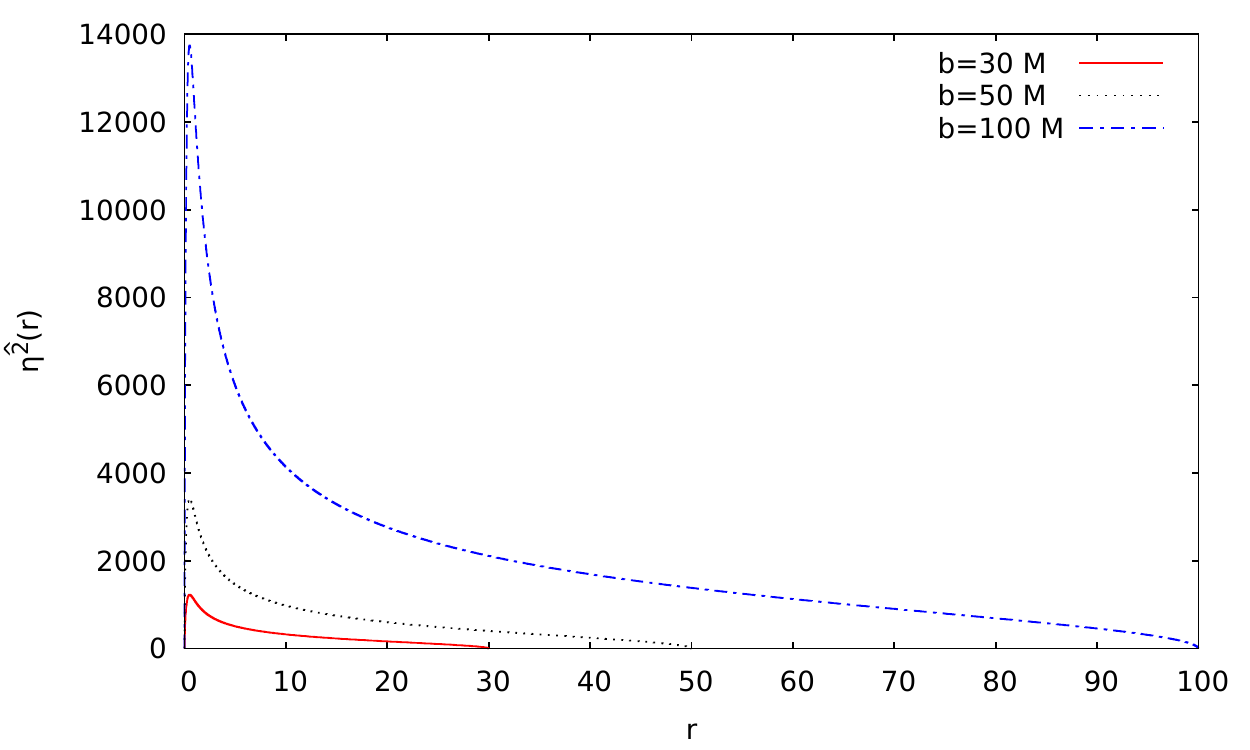}
\caption{The $\eta^{\hat{2}}$ component of the deviation vector, for different choices of $b$. In this figure, we have chosen $a=0.5\,M$, with initial conditions IC2, described in Eqs.~\eqref{IC21} and \eqref{IC22}. }
\label{eta2IC2acons}
\end{figure}
\end{center}

For the $\eta^{\hat{2}}$ component, 
the solutions with the IC2 are plotted in Figs.~\ref{eta2IC2bcons} and \ref{eta2IC2acons}.

In Fig.~\ref{eta2IC2bcons}, we let $b=100\,M$, $\frac{d\eta^{\hat{2}}}{d\tau}\mid_{r=b}=1$, 
 $M=1$ and plot the $\eta^{\hat{2}}$ component of the deviation vector 
 for various choices of the rotation parameter $a$ of the Kerr black hole. It can be seen from Fig.~\ref{eta2IC2bcons} that the behavior of $\eta^{\hat{2}}$ is essentially the same for different values of $a$ at large $r$. Besides that, for the Schwarzschild case ($a=0$) the component $\eta^{\hat{2}}$ goes to infinity as the body approaches the origin of the radial coordinate. For $a \neq 0$, the component 
$\eta^{\hat{2}}$ initially increases, reaches a maximum value and start decreasing due to the tidal force 
changing from stretching to compressing at $r=\sqrt{3}\,a$, similarly to the behavior presented in 
Fig.~\ref{eta2IC1bcons}.   The numerical result indicates that the deviation vector vanishes at 
$r=R^{stop}$, which is confirmed by the analytic solution presented in the next section.  
In Fig.~\ref{eta2IC2acons} we let  $a=0.5\,M$, $\frac{d\eta^{\hat{2}}}{d\tau}\mid_{r=b}=1$, 
$M=1$ and plot the $\eta^{\hat{2}}$ component of the deviation vector  
for various choices of $b$. The behavior of $\eta^{\hat{2}}$ is essentially the same as in 
Fig.~\ref{eta2IC2bcons}, and as we decrease the value of $b$, the maximum value of $\eta^{\hat{2}}$ also decreases.

\section{Analytic solutions to the geodesic deviation equations} \label{sec:analytic_solution}

In this section we find the components $\eta^{\hat{2}}$ and $\eta^{\hat{i}}$, $i=1,3$, 
of the geodesic deviation equations
analytically by varying the geodesic equations \eqref{cartereq0}-\eqref{cartereq3}.

\subsection{Radial geodesic deviation}
For the radial component $\eta^{\hat{2}}$ we can work with Eq.~\eqref{carteraximotion1}.  For a particle released from
rest at $r=b$ we find by letting $r=\dot{r}=0$ in Eq.~\eqref{carteraximotion1},
\begin{equation}
E^2 = 1 - \frac{2Mb}{b^2+a^2}.
\end{equation}
Then, Eq.~\eqref{carteraximotion1} reads
\begin{equation}
\label{f}\dot{r}=-\sqrt{2M}f(r,b),
\end{equation}
where
\begin{equation}
\label{3} f(r,b) = \left( \frac{r}{r^2+a^2} - \frac{b}{b^2+a^2}\right)^{\frac{1}{2}}.
\end{equation} 
Let a geodesic be perturbed in the radial direction and have the radial coordinate shifted to 
$r+\delta r$ and the parameter
$b$ to $b+\delta b$.  Then by varying Eq.~\eqref{f} and using the relation $d/d\tau = - \sqrt{2M}\,f(r,b)d/dr$, 
we obtain
\begin{eqnarray}
f\,\frac{d\,\delta r}{d\,r}=\frac{\partial f}{\partial r}\,\delta r+\frac{\partial f}{\partial b}\,\delta\,b,
\end{eqnarray}
which can be written as
\begin{equation}
\label{dr2}\frac{d}{d\,r}\left[\frac{\delta\,r}{f}\right]=\frac{1}{f^2}\frac{\partial f}{\partial b}\,\delta\,b.
\end{equation}
Solving 
this equation we find that 
\begin{eqnarray}
\label{drsol}\delta\,r=\delta b\,f\,\int\frac{1}{f^2}\frac{\partial f}{\partial b}\,dr+C\,f,
\end{eqnarray}
where $C$ is a constant. Thus, the variation $\delta r$ of an on-axis geodesic in the radial direction is given by
\begin{equation}
\delta r = C_1 \delta r^{(1)} + C_2 \widetilde{\delta r}^{(2)},
\end{equation}
where
\begin{eqnarray}
\delta r^{(1)} & = & f(r,b) \nonumber \\
& = & \left( \frac{r}{r^2+a^2} - \frac{b}{b^2+a^2}\right)^{\frac{1}{2}},\\
\widetilde{\delta r}^{(2)} & \propto &
f(r,b) \int\frac{1}{\left[ f(r,b)\right]^2}\frac{\partial f}{\partial b}\,dr \nonumber \\
& \propto & f(r,b)\int \frac{dr}{\left[ f(r,b)\right]^{3}}.
\end{eqnarray}

It is convenient for later purposes to rewrite the second independent variation by writing
\begin{eqnarray}
&& \left( \frac{r}{r^2+a^2} - \frac{b}{b^2+a^2}\right)^{-\frac{3}{2}} \nonumber \\
&& = \left( b+ \frac{a^2}{b}\right)^{\frac{3}{2}}\left[ \frac{r^2+a^2}{(r-a^2/b)(b-r)}\right]^{\frac{3}{2}},
\end{eqnarray}
and integrating by parts, using
\begin{equation}
\int\frac{dr}{(r-\alpha)^{\frac{3}{2}}(\beta - r)^{\frac{3}{2}}}
= \frac{2}{(\alpha-\beta)^2}\left( \sqrt{\frac{r-\alpha}{\beta-r}} - \sqrt{\frac{\beta-r}{r-\alpha}}\right).
\end{equation}
In this manner we find that the second independent variation can be chosen as
\begin{eqnarray}
\delta r^{(2)} & = &  (r^2+a^2)\left(2r- b- \frac{a^2}{b}\right) \nonumber \\
&& + 3f(r,b)
\int_r^b \dfrac{r'\left( 2r' - b - \frac{a^2}{b}\right)}{f(r',b)}\,dr'.
\end{eqnarray}

We find that both $\delta r^{(1)}$ and $\delta r^{(2)}$ satisfy the geodesic deviation equation~\eqref{eqdift}
for $\eta^{\hat{2}}$.  However, these do not describe geodesic deviation but the deviation of the radial coordinate.
Therefore, there is no \textit{a priori} reason why they satisfy the geodesic deviation equation, but
in fact $\delta r$ is  proportional to the deviation vector as we now demonstrate.  

Using the fact that the
deviation vector $(\delta t,\delta r)$ in the radial direction is orthogonal to the tangent vector to the geodesic,
$(\dot{t},\dot{r})$, and using the expressions for $\dot{t}$ and $\dot{r}^2$ in 
Eqs.~\eqref{carteraximotion0} and \eqref{carteraximotion1}, respectively, we find
\begin{equation}
\delta t = \frac{\Sigma_0}{E\Delta}\left( E^2 - \frac{\Delta}{\Sigma_0}\right)^{\frac{1}{2}}\delta r.
\end{equation}
Then, the square of the radial component of the deviation vector is
\begin{eqnarray}
(\eta^{\hat{2}})^2 & = & g_{\mu\nu}\delta x^\mu \delta x^\nu \nonumber \\
& = & (\delta r)^2/E^2,
\end{eqnarray}
where  $g_{\mu\nu}$ is the metric~\eqref{Kerr-metric} with $\theta=0$.  Hence
$\eta^{\hat{2}} = \delta r/E$.  This explains why $\delta r$ satisfies the geodesic deviation equation for 
$\eta^{\hat{2}}$.

The radial component of the deviation vector satisfying the IC1 is proportional to 
$\delta r^{(2)}$ and is given by
\begin{eqnarray}
\eta^{\hat{2}}_{(1)} & = & \frac{b}{(b^2+a^2)(b^2-a^2)}
\left[ (r^2+a^2)\left( 2r-b - \frac{a^2}{b}\right) \right. \nonumber \\
&& \left. \ \ \ \ \ + 3f(r,b)
\int_r^b \dfrac{r'\left(2r'-b-\frac{a^2}{b}\right)}{f(r',b)}\,dr'\right],
\end{eqnarray}
whereas that satisfying the IC2 is proportional to $\delta r^{(1)}$ and is given by
\begin{equation}
\eta^{\hat{2}}_{(2)} = \frac{2(b^2+a^2)^2}{\sqrt{2M}(b^2-a^2)}f(r,b).
\end{equation}
We note that $\eta^{\hat{2}}_{(1)} < 0$ and $\eta^{\hat{2}}_{(2)} = 0$ at $r=R^{stop}$.

\subsection{Angular geodesic deviation}  

The angular  deviation vectors can be found by examining the geodesic equation~\eqref{cartereq2} for small 
$\theta$.  By letting $K = a^2+\delta K$ and $\Phi=0$, we find from this equation
\begin{eqnarray}
\label{dtheta}\dot{\theta}^2=\frac{\delta K+\left(1-E^2\right)\,a^2\,\theta^2}{\left(r^2+a^2\right)^2},
\end{eqnarray}
to second order in $\theta$.
Using that
\begin{equation}
\frac{d\theta}{d\tau}=\dot{r}\,\frac{d\theta}{d\,r}= - \sqrt{2\,M}f(r,b)\frac{d\theta}{d\,r},
\end{equation}
we find from Eq.~\eqref{dtheta}
\begin{equation}
\label{dtheta2} 
\left( \frac{d\theta}{d\,r}\right)^2 = 
\frac{\delta K + (1-E^2)a^2\theta^2}{2M(r^2+a^2)^2\left[f(r,b)\right]^2}.
\end{equation}
The solutions $\theta(r)$ to this equation are given by\footnote{Equation \eqref{dtheta2} admits a constant solution
$\theta(r) = \theta_0 = \sqrt{|\delta K|/[(1-E^2)a^2]}$ if $\delta K < 0$.  
However, this is a spurious solution, and the function $\sqrt{r^2+a^2}\,\theta_0$
does not satisfy the geodesic deviation equation~\eqref{eqdifr}.}
\begin{equation}
\theta(r) = C_{+}\exp\left( \int_b^r\, F(r')\,dr'\right) + C_{-}\exp\left( - \int_b^r\,F(r')\,dr'\right),
\end{equation}
where
\begin{eqnarray}
F(r)=\left(\frac{b}{b^2+a^2}\right)^\frac{1}{2}\frac{a}{\left(r^2+a^2\right)f(r,b)},
\end{eqnarray}
and
\begin{equation}
C_{+}C_{-} =- \frac{(b^2+a^2)\delta K}{8Mba^2}.
\end{equation}
We have used the relation $1-E^2 = 2Mb/(b^2+a^2)$.

Since the metric takes the form $ds^2 = \cdots +\Sigma_0\,d\theta^2+\cdots$ on the symmetry axis, we expect that
the function $\eta^{\hat{i}} = \Sigma_0^{\frac{1}{2}} \theta(r)$ 
satisfies the geodesic deviation equation \eqref{eqdifr}.
This can readily be verified.  The angular component of the deviation vector satisfying the IC1 is
\begin{equation}
\eta^{\hat{i}}_{(1)} = \left( \frac{r^2+a^2}{b^2+a^2}\right)^{\frac{1}{2}}\cosh\left( \int_r^b F(r')\,dr'\right),
\end{equation}
whereas that satisfying the IC2 is
\begin{equation}
\eta^{\hat{i}}_{(2)} = \frac{(b^2+a^2)(r^2+a^2)^{\frac{1}{2}}}{\sqrt{2\,Mb}\,a}\sinh
\left( \int_r^b F(r')\,dr'\right).
\end{equation}

%%%%%%%%%%%%%%%%%%%%%%%%%%%%%%%%%%%%%%%%%%%%%%%%%%%%%%%%%%%%%%%%%%%%%%%%%%%%%%%%%%%%%%%%%%%%%%%%%%%%%%%%%%%%%%%%%
\section{Conclusion}
\label{Conclusion} 
%%%%%%%%%%%%%%%%%%%%%%%%%%%%%%%%%%%%%%%%%%%%%%%%%%%%%%%%%%%%%%%%%%%%%%%%%%%%%%%%%%%%%%%%%%%%%%%%%%%%%%%%%%%%%%
In this paper we studied tidal forces in Kerr spacetime for geodesic motion along the BH rotation axis. We used the equations of geodesic motion derived from the full integrability property of the geodesic equations in Kerr spacetime 
and took a suitable limit in Boyer-Lindquist coordinates though they are singular on this axis.  We used the
Carter tetrad basis attached to a body following the axial geodesic motion
in this limit. The results for the tidal forces were found, and we analyzed their dependence on the BH mass and rotation parameter. We noted that tidal forces in Kerr spacetime may vanish and change sign along the rotation axis infall. The point where the tidal force vanishes  is located outside the black hole event
horizon for rotation parameters greater than $a=\frac{\sqrt{3}}{2}M$. 
We pointed out
that the tidal forces and the Gaussian curvature at the event horizon vanish for the same value of the rotation parameter, i.e., $a=\frac{\sqrt{3}}{2}M$. We explained this fact by showing that the tidal forces and the Gaussian curvature are equal at the event horizon along the axis of symmetry for any value of $a$. 
%Therefore, when $a>\frac{\sqrt{3}}{2}M$ the tidal forces and the Gaussian curvature change signs at the event horizon. 
The implication of the change of sign of the Gaussian curvature is well known in the literature.
 It makes it impossible to embed 
the event horizon surface globally in a three dimensional Euclidean space. In order to analyze the 
physical implication of the change of sign of the tidal tensor, we solved the geodesic deviation equation, and compared the results with the Schwarzschild black hole case.

We have noted that tidal forces may cause compression or stretching, depending on the rotation parameter and the value of radial coordinate. We point out that tidal forces remain finite  except for the Schwarzschild case ($a=0$)
for which tidal forces diverge at $r=0$.

We analyzed the differential equations for the deviation vector associated to a geodesic motion along the symmetry axis of Kerr spacetime, and solved them  numerically and analytically 
as a function of the radial
coordinate, in order to analyze the effects of the changing of sign of the tidal tensor. We have chosen two types of initial conditions for the solutions of the geodesic deviation equation. The first solution represents a body constituted of dust with no initial internal motion at a point $b$ of the radial coordinate. The second initial condition  corresponds to letting 
such a body "explode" at a point $b$ of the radial coordinate. For values of the radial coordinate far away from the event horizon, the behavior of the deviation vector is qualitatively the same in Kerr and Schwarzschild spacetimes as expected. However, as the body approaches the event horizon of a Kerr BH, the behavior of the deviation vector may differ considerably from the Schwarzschild case because of the sign change in the
tidal forces.
%%%%%%%%%%%%%%%%%%%%%%%%%%%%%%%%%%%%%%%%%%%%%%%%%%%%%%%%%%%%%%%%%%%%%%%%%%%%%%%%%%%%%%%%%%%%%%%%%%%%%%%%%%%%%%%%%%%
\begin{acknowledgements}
We thank Carlos Herdeiro for useful discussions. The authors thank Funda\c{c}\~ao Amaz\^onia de Amparo a Estudos e Pesquisas (FAPESPA),  Conselho Nacional de Desenvolvimento Cient\'ifico e Tecnol\'ogico (CNPq) and Coordena\c{c}\~ao de Aperfei\c{c}oamento de Pessoal de N\'{\i}vel Superior (Capes) - Finance Code 001, for partial financial support.
This research has also received funding from the European Union's Horizon 2020 research and innovation programme under the H2020-MSCA-RISE-2017 Grant No. FunFiCO-777740. We also acknowledge the support from the Abdus Salam International Centre for Theoretical Physics through Visiting Scholar/Consultant Programme. 
One of the authors (A.\ H.)\  also thanks Federal University of Par\'a for kind hospitality.
\end{acknowledgements}

\appendix

\section{Parallel-propagated tetrad basis and the tidal tensor along any geodesic in Kerr spacetime}
\label{Marck_tetrad}
%%%%%%%%%%%%%%%%%%%%%%%%%%%%%%%%%%%%%%%%%%%%%%%%%%%%%%%%%%%%%%%%%%%%%%%%%%%%%%%%%%%%%%%%%%%%%%%%%%%%%%%%%%%%
In this Appendix we present the tetrad basis which is orthonormal and parallel-propagated along any geodesic in Kerr spacetime and the tidal tensor in this basis, summarizing some results of 
Ref. \cite{Marck:1983}. It is useful to consider the Carter's tetrad basis, which is given by \cite{B.Carter}
\begin{eqnarray}
\label{carter0}&&e^\mu_{(0)}=\left(\frac{(r^2+a^2)}{(\Sigma\,\Delta)^\frac{1}{2}},\,0\,,0\,,\,\frac{a}{(\Sigma\,\Delta)^\frac{1}{2}} \right),\\
&&e^\mu_{(1)}=\left(0,\,\left(\frac{\Delta}{\Sigma} \right)^\frac{1}{2},\,0,\,0\right),\\
\label{carter2}&&e^\mu_{(2)}=\left(0,\,0,\frac{1}{\Sigma^\frac{1}{2}},\,0 \right),\\
\label{carter3}&&e^\mu_{(3)}=\left(\frac{-a\,\sin\theta}{\Sigma^\frac{1}{2}},\,0,\,0,\,-\frac{(\sin\theta)^{-1}}{\Sigma^\frac{1}{2}} \right).
\end{eqnarray}
The dual basis is
\begin{eqnarray}
&&e^{(0)}_{\mu}dx^\mu=\left(\frac{\Delta}{\Sigma}\right)^\frac{1}{2}\left(dt-a\,\sin^2\theta\,d\phi\right),\\
&&e^{(1)}_{\mu}dx^\mu=\left(\frac{\Sigma}{\Delta} \right)^\frac{1}{2}\,dr,\\
&&e^{(2)}_{\mu}dx^\mu=\Sigma^\frac{1}{2}\,d\theta,\\
\label{inversecarter3}&&e^{(3)}_{\mu}dx^\mu=\left(\frac{\sin\theta}{\Sigma^\frac{1}{2}}\right)\left(a\,dt-(r^2+a^2)
\,d\phi\right).
\end{eqnarray}
We note that, in the limit $\theta\to 0$ with $\theta>0$, the vectors $e^\mu_{(0)}$, $e^\mu_{(1)}$, $e^\mu_{(2)}$ 
and $- e^\mu_{(3)}$ point in the directions of increasing $t$, $r$, $\theta$ and $\phi$, respectively.  As a result, the
Carter tetrad on the geodesic along the rotation axis rotates with the angular velocity given by $\dot{\phi}$ in 
Eq.~\eqref{carteraximotion3} relative to the constant-$\phi$ hypersurfaces.

 We will write the components of the parallel-propagated tetrad basis in terms of the Carter tetrad. The first vector of the parallel-propagated tetrad basis along any geodesic  may be chosen to be the components of the 4-velocity given by Eqs.~\eqref{cartereq0}-\eqref{cartereq3}, 
i.e.\ 
\begin{equation}
\label{l0}
\lambda^\mu_{\ \hat{0}}=\dot{x}^\mu,
\end{equation}
which is a timelike vector.
We can also write the components of the vector \eqref{l0} 
in the Carter's tetrad basis through
\begin{equation}
\label{marcktetrad0}\lambda^{(a)}_{\ \hat{0}}=\dot{x}^\mu\,e^{(a)}_{\mu},
\end{equation}
so that
 \begin{eqnarray}
\label{ftv0}&& \lambda^{(0)}_{\ \hat{0}}=\frac{1}{(\Delta\,\Sigma)^\frac{1}{2}}\left(E(r^2+a^2)-a\,\Phi \right),\\
&& \lambda^{(1)}_{\ \hat{0}}=\left(\frac{\Sigma}{\Delta}\right)^\frac{1}{2}\,\dot{r},\\
&& \lambda^{(2)}_{\ \hat{0}}=\Sigma^\frac{1}{2}\,\dot{\theta},\\
 \label{ftv3}&& \lambda^{(3)}_{\ \hat{0}}=\frac{1}{\Sigma^\frac{1}{2}}\left(a\,E\,\sin\theta-\frac{\Phi}{\sin\theta} \right).
 \end{eqnarray}
 
The second vector of the tetrad basis may be found with the aid of the Killing-Yano tensor $f_{\mu\,\nu}$ for the Kerr geometry that is anti-symmetric and satisfies \cite{Ref-18}
 \begin{equation}
 \label{yanoeqsim}\nabla_\mu\,f_{\nu\,\beta}+\nabla_\nu\,f_{\mu\,\beta}=0,
 \end{equation} 
 which is given by
 \begin{eqnarray}
\nonumber\frac{1}{2}f_{\mu\,\nu}dx^\mu\wedge dx^\nu=a\,\cos\theta\,dr \wedge (dt-a\,\sin^2\theta\, d\phi)\\
\label{yanotensor}+r\,\sin\theta\,d\theta \wedge (-a\,dt+(r^2+a^2)\,d\phi).
\end{eqnarray}
As pointed out by  Penrose \cite{Ref-18}, the unit vector
\begin{equation}
 \label{Killing-Yanovec}L^\mu=K^{-\frac{1}{2}}f^\mu_{\  \ \nu}\dot{x}^\nu,
 \end{equation}
 is parallel-propagated along any geodesic and is orthogonal to $\dot{x}^\mu$.  Thus,
we may choose $L^\mu$ to be the second vector of the tetrad basis. Its components, in Carter's tetrad basis, 
are given by
  \begin{eqnarray}
\label{stv0} &&\lambda^{(0)}_{\ \hat{2}}=\left(\frac{\Sigma}{K\,\Delta} \right)^\frac{1}{2}\,a\,\cos\theta\,\dot{r},\\
&&\lambda^{(1)}_{\ \hat{2}}=\frac{1}{(K\Sigma\,\Delta)^\frac{1}{2}}\,a\,\cos\theta\left(E(r^2+a^2)-a\,\Phi\right)\\
&&\lambda^{(2)}_{\ \hat{2}}=-\frac{1}{(K\Sigma)^\frac{1}{2}}\,r\,\left(a\,E\,\sin\theta-\frac{\Phi}{\sin\theta} \right),\\
\label{stv3}&&\lambda^{(3)}_{\ \hat{2}}=\left(\frac{\Sigma}{K}\right)^\frac{1}{2}\,r\,\dot{\theta}.
 \end{eqnarray}
The other two vectors chosen by Marck~\cite{Marck:1983} to complete the tetrad basis are given in components as
  \begin{eqnarray}
 \label{tildetf10}&&\tilde{\lambda}^{(0)}_{\hat{1}}=\Lambda\left(\frac{\Sigma}{K\,\Delta} \right)^\frac{1}{2}\,r\,\dot{r},\\
 &&\tilde{\lambda}^{(1)}_{\hat{1}}=\Lambda\left(\frac{1}{K\,\Sigma\,\Delta}\right)^\frac{1}{2}\,r\,\left(E(r^2+a^2)-a\,\Phi \right),\\
 &&\tilde{\lambda}^{(2)}_{\hat{1}}=\frac{1}{\Lambda}\left(\frac{1}{K\,\Sigma} \right)^\frac{1}{2}\,a\,\cos\theta\left(a\,E\,\sin\theta-\frac{\Phi}{\sin\theta}\right),\\
 \label{tildetf13}&&\tilde{\lambda}^{(3)}_{\hat{1}}=-\frac{1}{\Lambda}\left(\frac{\Sigma}{K}\right)^\frac{1}{2}\,a\,\cos\theta\,\dot{\theta},
  \end{eqnarray}
  and
  \begin{eqnarray}
\label{tildetf30}&&\tilde{\lambda}^{(0)}_{\hat{3}}=\Lambda\,\left(\frac{1}{\Sigma\,\Delta}\right)^\frac{1}{2}\left(E\left(r^2+a^2\right)-a\,\Phi\right),\\
&&\tilde{\lambda}^{(1)}_{\hat{3}}=\Lambda\left(\frac{\Sigma}{\Delta}\right)^\frac{1}{2}\,\dot{r},\\
&&\tilde{\lambda}^{(2)}_{\hat{3}}=\frac{1}{\Lambda}\,\Sigma^\frac{1}{2}\,\dot{\theta},\\
\label{tildetf33}&&\tilde{\lambda}^{(3)}_{\hat{3}}=\frac{1}{\Lambda}\left(\frac{1}{\Sigma}\right)^\frac{1}{2}\left(a\,E\,\sin\theta-\frac{\Phi}{\sin\theta}\right),
\end{eqnarray}  
where
\begin{equation}
\Lambda=\sqrt{\frac{K-a^2\,\cos^2\theta}{K+r^2}}.
\end{equation} 

 Although the vectors 
$\tilde{\lambda}^{(a)}_{\hat{1}}$ and $\tilde{\lambda}^{(a)}_{\hat{3}}$ are orthogonal to 
$\lambda^{(a)}_{\ \hat{2}}$  and $\lambda^{(a)}_{\ \hat{0}}$, they are not parallel-propagated along 
the geodesic in general.  
In order to find orthonormal vectors that are parallel-propagated along the geodesic, one 
performs a time-dependent spatial rotation of $\tilde{\lambda}^{(a)}_{\hat{1}}$ and 
$\tilde{\lambda}^{(a)}_{\hat{3}}$: 
\begin{eqnarray}
\label{ttv}&&\lambda^{\mu}_{\hat{1}}=\tilde{\lambda}^{\mu}_{\hat{1}}\,\cos\Psi-\tilde{\lambda}^{\mu}_{\hat{3}}\,\sin\Psi,\\
\label{fotv}&&\lambda^{\mu}_{\hat{3}}=\tilde{\lambda}^{\mu}_{\hat{1}}\,\sin\Psi+\tilde{\lambda}^{\mu}_{\hat{3}}\,\cos\Psi.
\end{eqnarray}
Requiring 
that the vectors $\lambda^{(a)}_{\hat{1}}$ and $\lambda^{(a)}_{\hat{3}}$ 
be parallel-propagated along the geodesic, i.e.:
\begin{equation}
\label{g}\lambda^\mu_{\hat{0}}\nabla_\mu\lambda^\nu_{\hat{i}}=0, \ \ \ \ \ \ \ i=(1,3),
\end{equation}
one finds that the proper-time derivative of $\Psi$ is given by 
\begin{equation}
\label{psi}\dot{\Psi}=\frac{K^\frac{1}{2}}{\Sigma}\left(\frac{E(r^2+a^2)-a\,\Phi}{K+r^2}+\frac{a(\Phi-a\,E\,\sin^2\theta)}{K-a^2\,\cos^2\theta}\right).
\end{equation}
The tetrad basis  $\left(\lambda^{(a)}_{\hat{0}}, \lambda^{(a)}_{\hat{1}} , \lambda^{(a)}_{\hat{2}},\lambda^{(a)}_{\hat{3}}\right)$ is orthonormal and parallel-propagated along the geodesic.

One can computes the tidal tensor in this tetrad basis using 
\begin{equation}
\label{tidaltensorincartercomp}K_{\hat{e}\,\hat{f}}=R_{(a)\,(b)\,(c)\,(d)}\lambda^{(a)}_{\hat{0}}\lambda^{(b)}_{\hat{e}}\lambda^{(c)}_{\hat{0}}\lambda^{(d)}_{\hat{f}},
 \end{equation}
  where $R_{(a)\,(b)\,(c)\,(d)}$ are the components of the Riemann tensor written in Carter's tetrad basis. 
The components of 
the tidal tensor \eqref{tidaltensorincartercomp}  are given by
  \begin{eqnarray}
 \nonumber&&K_{\hat{1} \hat{1}}=\left[1-3\,S\,T\frac{\left(r^2-a^2\,\cos^2\theta\right)}{K\,\Sigma^2}\cos^2\Psi\right]\,I_1\\
 \label{Tidaltensor11}&&+6\,a\,r\,\cos\theta\,\cos^2\Psi\frac{S\,T}{K\,\Sigma^2}I_2,\\
  \nonumber&&K_{\hat{1} \hat{2}}=\frac{3\,(S\,T)^\frac{1}{2}}{K\,\Sigma^2}\cos\Psi\left[(a^2\,\cos^2\theta\,S-r^2\,T)I_2\right.\\
 \label{Tidaltensor12}&&\left. -a\,r\,\cos\theta(S+T)I_1 \right],\\
 \nonumber &&K_{\hat{1} \hat{3}}=\frac{3\,S\,T}{K\,\Sigma^2}\,\cos\Psi\,\sin\Psi\left[(a^2\,\cos^2\theta-r^2)I_1\right.\\
  \label{Tidaltensor13}&&\left.+2\,a\,r\,\cos\theta\,I_2\right],\\
\nonumber &&K_{\hat{2} \hat{2}}=\left(1+3\,\frac{r^2 \,T^2-a^2\,\cos^2\theta\,S^2}{K\,\Sigma^2}\right)I_1\\
 \label{Tidaltensor22} &&-6\,a\,r\,\cos\theta\frac{S\,T}{K\,\Sigma^2}I_2,\\
  \nonumber&&K_{\hat{2} \hat{3}}=\frac{3\,(S\,T)^\frac{1}{2}}{K\,\Sigma^2}\sin\Psi\left[(a^2\,\cos^2\theta\,S-r^2\,T)I_2\right.\\
 \label{Tidaltensor23}&&\left. -a\,r\,\cos\theta(S+T)\,I_1\right],\\
 \nonumber &&K_{\hat{3} \hat{3}}=\left[1-\frac{3\,S\,T\,(r^2-a^2\,\cos^2\theta)}{K\,\Sigma^2}\,\sin^2\Psi\right]
I_1\\
\label{Tidaltensor33} &&+6\,a\,r\,\cos\theta\frac{S\,T}{K\,\Sigma^2}\sin^2\Psi\,I_2,
 \end{eqnarray}
 where
 \begin{eqnarray}
 &&S=r^2+K,\\
 &&\label{Ttidalforce}T=K-a^2\,\cos^2\theta,\\
 &&\label{I1}I_1=\frac{M\,r}{\Sigma^3}(r^2-3\,a^2\,\cos^2\theta),\\
 &&\label{I2tidalforce}I_2=\frac{M\,a\,\cos\theta}{\Sigma^3}(3\,r^2-a^2\,\cos^2\theta).
 \end{eqnarray}
%%%%%%%%%%%%%%%%%%%%%%%%%%%%%%%%%%%%%%%%%%%%%%%%%%%%%%%%%%%%%%%%%%%%%%%%%%%%%%%%%%%%%%%%%%%%%%%%%%%%%%%%%%%%%%%%%%

\end{document}